\documentclass[11pt,a4paper]{article}
\pdfoutput=1
\usepackage[pdftex]{graphics}
\usepackage{jheppub}
\usepackage{amsmath}
\usepackage{amssymb}
\usepackage{graphicx}
\usepackage{slashed}
\usepackage{booktabs}
\usepackage{dsfont}
\newcommand{\nn}{\nonumber}

\def\half{\frac{1}{2}}

\def\det{{\rm det}}

\def\CA{{\cal A}}

\def\CF{{\cal F}}

\def\CH{{\cal H}}
\def\CI{{\cal I}}
\def\CJ{{\cal J}}
\def\CK{{\cal K}}
\def\CL{{\cal L}}
\def\CM{{\cal M}}
\def\CN{{\cal N}}

\def\CP{{\cal P}}

\def\CT{{\cal T}}

\def\CW{{\cal W}}

\def\a{\alpha}

\def\g{\gamma}
\def\d{\delta}

\def\G{\Gamma}

\def\sc{\mathsf{c}}

\def\sst#1{{\scriptscriptstyle #1}}
\def\0{{\sst{(0)}}}
\def\1{{\sst{(1)}}}
\def\2{{\sst{(2)}}}
\def\3{{\sst{(3)}}}
\def\4{{\sst{(4)}}}
\def\5{{\sst{(5)}}}
\def\6{{\sst{(6)}}}
\def\7{{\sst{(7)}}}
\def\8{{\sst{(8)}}}

\def\ft#1#2{{\textstyle{\frac{{\scriptstyle #1}}{{\scriptstyle #2}}}}}
\def\fft#1#2{{\frac{#1}{#2}}}
\newcommand{\be}{\begin{equation}}
\newcommand{\ee}{\end{equation}}
\newcommand{\bea}{\begin{eqnarray}}
\newcommand{\eea}{\end{eqnarray}}

\newcommand{\ep}{\epsilon}

\title{Holography of 3d-3d correspondence at Large N}

\author[a]{Dongmin Gang,}
\author[b]{Nakwoo Kim,}
\author[a,c]{Sangmin Lee,}

\affiliation[a]{School of Physics, Korea Institute for Advanced Study, 
\\
85 Hoegiro Dongdaemun-gu, Seoul 130-722, Korea}
\affiliation[b]{Department of Physics and Research Institute of Basic Science,
Kyung Hee University, 
\\
26 Kyungheedaero Dongdaemun-gu, Seoul 130-701, Korea}
\affiliation[c]{Center for Theoretical Physics, Department of Physics and Astronomy, College of Liberal Studies, 
Seoul National University, 1 Gwanakro Gwanak-gu, Seoul 151-742, Korea}

\emailAdd{arima275@kias.re.kr}
\emailAdd{nkim@khu.ac.kr}
\emailAdd{sangmin@snu.ac.kr}

\abstract{  
We study the physics of multiple M5-branes compactified on a hyperbolic 3-manifold. On the one hand, it leads to the 3d-3d correspondence 
which maps an $\mathcal{N}=2$ superconformal field theory to 
a pure Chern-Simons theory on the 3-manifold. 
On the other hand, it leads to a warped AdS$_4$ geometry in M-theory 
holographically dual to the superconformal field theory. 
Combining the holographic duality and the 3d-3d correspondence, we propose a conjecture for the large $N$ limit of the perturbative free energy of a Chern-Simons theory on hyperbolic 3-manifold. 
The conjecture claims that the tree, one-loop and two-loop terms all share the same $N^3$ scaling behavior and are proportional to the volume of the 3-manifold, while the three-loop 
and higher terms are suppressed at large $N$. Under mild assumptions, 
we prove the tree and one-loop parts of the conjecture. For the two-loop part, we test the conjecture numerically in a number of examples 
and find precise agreement. We also confirm the suppression of higher loop terms in a few examples. 
}

\begin{document}
\maketitle

\section{Introduction} \label{sec : introduction}
Although M5-brane is one of the most fundamental objects in M-theory, the physics of multiple M5-branes still remains mysterious. 
For a single M5 brane, the low energy effective world-volume theory is a free theory of an Abelian self-dual 2-form tensor multiplet with known Lagrangian description.
The  world-volume theory for $N \geq 2$ coincident M5-branes is called 6d $A_{N-1}$ (2,0) theory. It has 6d $(2,0)$ superconformal symmetry whose bosonic subgroup is $SO(2,6)\times SO(5)_R$. The (2,0) theory is expected to be a kind of non-Abelian tensor theory, but the attempts 
to write down a Lagrangian have not yet reached a stage 
where all quantum observables of the theory 
can be computed straightforwardly, at least in principle, 
from the Lagrangian. Alternative approaches to study the (2,0) theory invoke dualities or topologically protected observables. One famous result is the $N^3$ scaling of the theory's degrees of freedom, which was argued 
to be true on the basis of holographic principle and  anomaly calculations \cite{Klebanov:1996un,Henningson:1998gx,Harvey:1998bx,Yi:2001bz}.

Recently much attention have been paid to  the lower dimensional theories obtained by compactifying the 6d  (2,0) theory on an internal manifold $M$ with a partial topological twisting along $M$. 
A large class of 4d superconformal field theories (SCFTs), 
called theories of class $S$ \cite{Gaiotto:2009hg,Gaiotto:2009we}, have 
been constructed with $M$ being Riemann surfaces with punctures. 
This type of constructions provide new ways to understand many aspects of lower dimensional theories from the geometry of $M$. In particular, S-dualities among theories of class $S$ correspond to different pants decompositions of a Riemann surface \cite{Gaiotto:2009we}. 
The geometric interpretation of S-dualities also led to the celebrated AGT conjecture \cite{Alday:2009aq,Wyllard:2009hg}, 
which states that some supersymmetric quantities such as the partition function (ptn) on a squashed sphere can be identified with the ptn for some bosonic theory on $M$.   
Furthermore, new examples of holographic AdS/CFT duality can be obtained from these constructions. The gravity duals of 4d theories class of $S$ were first studied in \cite{Gaiotto:2009gz}.  

Another important advantage of this approach is that we can learn something about the 6d theory by studying the lower dimensional theories.
% which could not be understood from conventional field-theoretic methods.  
One  example is the calculation of the superconformal index for the 6d theory from 5d maximally supersymmetric Yang-Mills theory regarded as an $S^1$-reduction of the 6d theory \cite{Kim:2012ava,Kim:2012qf,Kim:2013nva}.  The famous $N^3$ behavior of the 6d theory can also be understood by calculating such physical quantities as anomaly coefficients in even dimensions \cite{Gaiotto:2009gz,Benini:2013cda} or sphere ptns in odd dimensions \cite{Kim:2012ava}.

In this paper, we study 3d SCFTs, called $T_{N}[M]$, constructed by compactifying the 6d $A_{N-1}$ (2,0) theory on a 3-manifold $M$. In the compactification, we perform a topological twisting along $M$ using an $SO(3)_R$ subgroup of the $SO(5)_R$ R-symmetry.  This twisting preserves a quarter of the supercharges and the 3d theories at IR fixed point have 3d $\CN=2$ superconformal symmetry.  These theories enjoy several dualities. 3d mirror symmetries can be interpreted as ambiguities in the choice of ideal triangulation of $M$ \cite{Dimofte:2011ju}. 
The 3d-3d correspondence identifies supersymmetric ptns of $T_{N}[M]$ on a curved background $B$ to certain topological invariants on $M$. 
For $B$ being a squashed three sphere $S^3_b$ \cite{Hama:2011ea}, 
the details of the correspondence were given in \cite{Terashima:2011qi,Dimofte:2011ju}. For $B= S^2\times S^1$, they were given in \cite{Dimofte:2011py,Gang:2013sqa}. In both cases, the topological invariants are CS ptns on $M$ with a complex gauge group and suitable CS levels which depend on $B$. 
Physical derivations of the dualities were given in \cite{Yagi:2013fda,Lee:2013ida,Cordova:2013cea} by studying the compactification  of the (2,0) theory on $B$.  

In the large $N$ limit, we can learn more about the 3d SCFTs  by considering the $AdS_4$ gravity duals. Ignoring subtle structures near the boundaries of internal manifold $M$, the 3d theory can be engineered by taking the IR fixed point of the world-volume theory for $N$ coincident M5-branes wrapping the $\mathbb{R}^{1,2}\times M$ subspace of $\mathbb{R}^{1,2}\times T^*M \times \mathbb{R}^2$ in eleven dimensions, where $T^*M$ is the cotangent bundle of $M$. Motivated by the brane configuration, the gravity duals for the 3d theories were studied in \cite{Gauntlett:2000ng} building upon earlier work \cite{Pernici:1984nw}.  The dual supergravity solution exists only when $M$ is hyperbolic, which might imply that for non-hyperbolic $M$, the IR fixed point is a topological theory without any physical degree of freedom.

Our main goal of the present paper is to combine the holographic duality 
and the 3d-3d correspondence to make a strong conjecture for 
the large $N$ behavior of the perturbative expansion of the CS theory on $M$. The statement of the conjecture and some preliminary evidences were announced earlier by the authors in \cite{Gang:2014qla}. In this paper, we give a more detailed 
account of the reasoning behind the conjecture and present more evidences, analytic and numerical, supporting the conjecture. 

The AdS/CFT dictionary states that the partition function of the 3d SCFT on $S^3_b$ is equal to the partition function of the gravity on a squashed Euclidean AdS$_4 (b)$ whose asymptotic boundary is $S^3_b$.  For  AdS$_4$/CFT$_3$ arising from multiple M2-branes, the equality has been extensively verified \cite{Drukker:2010nc,Herzog:2010hf,Martelli:2011qj,Cheon:2011vi,Jafferis:2011zi}. 
 The large $N$ limit of the CFT corresponds to the classical limit of the gravity. The free energy, $\mathcal{F}_b = - \log |Z_{S^3_b}|$, in this limit is the holographically renormalized on-shell action on the gravity side. Using the supergravity solution in  \cite{Gauntlett:2000ng}, the free energy $\CF^{\textrm{gravity}}$ will be calculated in section  \ref{sec : sugra}. The result is summarized in eq.~\eqref{gravity free energy}. On the SCFT side, instead of computing the free energy 
 from of $T_N[M]$ directly, we invoke the 3d-3d correspondence which states $Z_{S^3_b}\left( T_{N}[M] \right) = Z^{\textrm{CS}}_{N}[M;\hbar]$, a $PGL(N)$ CS ptn  on $M$ with a coupling constant $\hbar:=2\pi i b^2$. 

Under two mild assumptions on  the topological invariant $Z^{{\rm CS}}_N(M;\hbar)$, stated in \eqref{Two things on CS ptn-1} and \eqref{Two things on CS ptn-2}, we  show that  the holographic prediction $\CF^{\textrm{gravity}} = -\log |Z^{{\rm CS}}_N(M;\hbar)|$ implies an interesting  large $N$  behavior  of  the perturbative invariants of the $PGL(N)$ CS theory on $M$. The result, summarized in \eqref{conjecture}, 
is the main conjecture of this paper.
Roughly speaking, the conjecture claims that the tree, one-loop and two-loop perturbative invariants $(S_0, S_1, S_2)$ all share the same $N^3$ scaling behavior, whereas 
all higher loop invariants $S_{n\ge 3}$ are relatively suppressed. 
An analytic proof of the conjecture for $S_0$ and $S_1$ are given in section  \ref{subsection : conjecture}. In section \ref{sec : dimofte-state integral}, we  give some numerical evidences for higher order invariants for various knot complements using Dimofte's state-integral model \cite{Dimofte:2011gm,Dimofte:2012qj}.  

The main conjecture has passed all analytic and numerical tests so far, 
which seems to suggest that the chain of dualities is consistent and that our assumptions on the CS invariants are valid.
We leave the complete proof of the whole conjecture as a future problem. We conclude with discussions on future directions in section \ref{sec : conclusion}.

%%%%%%%%%%%%%%%%%%%%%%%%%%%%%%%
\section{Supergravity analysis} \label{sec : sugra}
In this section, we review the gravity dual of $T_{N}[M]$ and calculate the gravitational free energy  $\mathcal{F}^{\rm gravity}$ in the supergravity approximation.
Combining the results from \cite{Gauntlett:2000ng,Martelli:2011fu},
one concludes that
\begin{align}
\CF^{\rm gravity}_b =
\frac{N^3}{12\pi} \left(b+\frac{1}{b}\right)^2\textrm{vol}(M) \;,
\label{gravity free energy}
\end{align}
with subleading corrections in $1/N$. This result can be computed from an effective $D=4$ gauged supergravity theory relevant to the current setup of M5-branes wrapped on a special Lagrangian
3-cycle in a Calabi-Yau three-fold.
The gravitational free energy is directly related to the $D=4$ gravitational
constant \cite{Emparan:1999pm}, which is in turn determined by the volume of the internal $D=7$ space and the overall length scale of the $D=11$ metric. The hyperbolic
space $M$ is a part of the internal space, thus a factor of ${\rm vol} (M)$ in \eqref{gravity free energy}. The $N^3$ dependence on the number of M5-branes generically appears
when we relate the overall length scale with $N$ using the four-form
flux quantization condition in M-theory.

In fact, however, the  solution in  \cite{Gauntlett:2000ng} cannot be the complete gravity dual of $T_{N}[M]$ when 3-manifold $M$ is a knot (or link) complement since there is no tunable parameters in the  solution which parameterize type of defects along knot (or link). When we say the 3d theory $T_{N}[M]$ is associated with a knot complement $M=S^3\backslash \CK$, we  need to specify what type of defects are placed along the knot. As an analogy, consider the 4d theories of class $S$ associated with a Riemann surface with  punctures. Depending on the type of defects at the punctures, the corresponding 4d SCFTs are different and have different dual supergravity geometry \cite{Gaiotto:2009gz}. As far as the leading $N^3$-terms of conformal anomaly coefficients $a$ and $c$ are concerned, however, the detailed structures of the supergravity solution associated to punctures are irrelevant if all punctures are `full' (or maximal) punctures. The leading $N^3$-terms only depend on the Euler character of the Riemann surface regardless of existence of punctures. In the same vein, we expect that, despite the incompleteness of the solution, the gravity free energy formula  \eqref{gravity free energy} is reliable even for knot complements $M$ as far as the $N^3$-term is concerned, if defects along the knots are  ``full knots''.  The full knot defects can be realized as $N$  M5-branes along  the unit co-normal bundle of a knot $\CK$ in $T^{*} S^3$ intersecting with $N$ M5-branes on $S^3$.

%%%%%%%%%%%%%%%%%%%%%%%
\subsection{$D=7$ maximal supergravity}
%%%%%%%%%%%%%%%%%%%%%%%

When one is to look for nontrivial M5-brane backgrounds in
the near-horizon limit, it is convenient to use the $D=7$ maximally gauged supergravity first and then uplift the
solution back to $D=11$. The $D=7$ theory contains, as bosonic degrees of freedom, the metric tensor,
$SO(5)$ gauge fields $A^{ij}$, 14 scalar fields constituting a symmetric, unimodular matrix $T_{ij}$ parametrizing the  coset $SL(5,\mathbb{R})/SO(5)$, and
5 three-form tensor fields $S^i_{(3)}$. We use $i,j=1,\ldots,5$ to represent $SO(5)$ indices, and subscripts in parentheses ($\omega_{(n)}$) to denote $n$-forms.
We will follow the notation in \cite{Donos:2010ax}.
For the original construction, the readers are referred to \cite{Pernici:1984nw}.

The Lagrangian as a seven-form is written as
\begin{align}
{\cal L}_{7} =&  \; R * {\bf 1} - \tfrac{1}{4} T^{-1}_{ij} * D T_{jk} \wedge T^{-1}_{kl} D T_{li}
-\tfrac{1}{4} T^{-1}_{ik} T^{-1}_{jl} * F^{ij}_{\2} \wedge F^{kl}_{\2}
\nn
\\
& -\tfrac{1}{2} T_{ij} * S^i_{\3} \wedge S^j_{\3} + \tfrac{1}{2g} S^i_{\3} \wedge D S^{i}_{\3}
- \tfrac{1}{8g} \epsilon_{ij_1 \cdots j_4} S^i_{\3} \wedge F^{j_1j_2}_{\2} \wedge F^{j_3j_4}_{\2}
\nn\\
& +\tfrac{1}{g} \Omega_{\7} - V * {\bf 1} .
\label{L7}
\end{align}
Here the covariant derivatives are defined as
\begin{align}
DT_{ij} &\equiv dT_{ij} + g A^{ik}_{\1} T_{kj} + g A^{jk}_{\1} T_{ik} ,
\\
D S^i_\3 & \equiv d S^i_\3 + g A^{ij}_\1 \wedge S^j_\3 ,
\\
F^{ij}_\2 &\equiv d A^{ij}_\1 + g A^{ik}_\1 \wedge A^{kj}_\1 .
\end{align}
The scalar potential $V$ is given by $T_{ij}$ as follows,
\begin{equation}
V = \ft12  g^2 \Big(2 T_{ij}\, T_{ij} - (T_{ii})^2 \Big)\, .
\end{equation}
The seven-form $\Omega_\7$ is a quartic Chern-Simons type term built from the $SO(5)$ Yang-Mills fields
and its explicit form will not concern us in this paper.

It is established that any solution of the above $D=7$ system gives rise to a solution of
$D=11$ supergravity \cite{Nastase:1999cb,Nastase:1999kf}. Using the notation of
\cite{Cvetic:2000ah}, the uplifting formula for metric is
\be
d s_{11}^2 = \Delta^{1/3}\, ds_{7}^2 + \frac 1{g^2}\Delta^{-2/3}\,
T^{-1}_{ij}\, D\mu^i\, D\mu^j\, .\label{metel}
\ee
And for the four-form field,
\begin{align}
G_\4 =& \;\frac{\Delta^{-2}}{4!g^3}\, \ep_{i_1\cdots i_5}\, \Big[
-  U\,  \mu^{i_1} D\mu^{i_2}\wedge D\mu^{i_3}\wedge D\mu^{i_4}\wedge
D\mu^{i_5}
\nn \\
& + 4 \, T^{i_1 m}\, DT^{i_2 n}\, \mu^m\, \mu^n\,
D\mu^{i_3}
\wedge D\mu^{i_4} \wedge D\mu^{i_5}+ 6g \Delta F_\2^{i_1 i_2} \wedge
D\mu^{i_3}\wedge D\mu^{i_4}\, T^{i_5 j}\, \mu^j \Big]
\nn \\
&- T_{ij}\,
{*S_\3^i}\, \mu^j + \fft1{g}\, S_\3^i \wedge D\mu^i\, .\label{4form}
\end{align}
Here $\mu^i, i=1,\ldots,5$ are angular coordinates for $S^4$, i.e., $\sum_i (\mu^i)^2=1$,
and
\begin{align}
U \equiv 2 T_{ij}\, T_{jk}\, \mu^i\, \mu^k - \Delta\, T_{ii}\,, \qquad
\Delta \equiv T_{ij}\, \mu^i\, \mu^j\,,\qquad
D\mu^i \equiv d\mu^i + g A_\1^{ij}\, \mu^j\, \,.
\end{align}

It is easily checked that the trivial $AdS_7$ vacuum of \eqref{L7} with vanishing form-fields and $T_{ij}=\delta_{ij}$
has radius $2/g$. Using then the above uplifting formula, in $D=11$ we have
\begin{align}
ds^2_{11} & = \frac{4}{g^2} ds^2(AdS_7) + \frac{1}{g^2}ds^2(S^4) \, ,
\label{ads7}
\\
G_\4 &= \frac{3}{g^3} {\rm vol}(S^4) \, .
\end{align}
Here both $ds^2(AdS_7)$ and $ds^2(S^4)$ are normalized to
have unit radius.

The standard convention for $D=11$ supergravity is to make the
 Planck length appear in the action as follows,
\be
{S} = \frac{1}{(2\pi)^8 l_P^9} \int \left(R *{\bf 1} - \frac{1}{2} G_\4 \wedge * G_\4 - \frac{1}{6} C_\3 \wedge
G_\4 \wedge G_\4 \right) \, .
\ee
The four-form flux quantization then gives the M5-brane number as
\be
N_{M5} = \frac{1}{(2\pi l_P)^3} \int_{X_4} G_\4 \, ,
\label{Q5}
\ee
where $X_4$ is a 4-cycle in $D=11$ spacetime. Using this relation, we may  rewrite
\eqref{ads7} as
\be
ds^2_{11} = l^2_P (\pi N)^{2/3}  \left[ 4 ds^2(AdS_7) +  ds^2(S^4) \right] \, ,
\ee
which is  for instance the same as Eq.(3.2) of \cite{Maldacena:1997re}.

%%%%%%%%%%%%%%%%%
\subsection{$AdS_4\times H^3$ Solution as wrapped M5-brane}
%%%%%%%%%%%%%%%%%
It is known that, in addition to the maximally supersymmetric $AdS_7$
solution, the action \eqref{L7} allows a variety
of supersymmetric magnetically charged solutions \cite{Pernici:1984nw}. In terms of
M-theory, such solutions are interpreted as M5-branes wrapped on supersymmetric
cycles \cite{Maldacena:2000mw,Gauntlett:2000ng}. Among many possibilities, we are particularly
interested in the case of M5-branes wrapped on a special Lagrangian 3-cycle within
a Calabi-Yau three-fold. To preserve supersymmetry, one twists the M5-brane theory
by coupling it to a subalgebra of R-symmetry, which is in this case $SO(3)\subset SO(5)_R$.
In the $D=7$ gauged supergravity,
this procedure is implemented by turning on a $SO(3)$ gauge field to cancel precisely the
contribution of the spin connection on the 3-cycle $\Sigma_3$. One finds $AdS_4\times \Sigma_3$
fixed point solutions when $\Sigma_3=H^3$.

A convenient way of solving the equation of motion, when we adopt the twisting prescription
above, is to consider dimensionally reduced effective Lagrangian \cite{Gauntlett:2002rv,Donos:2010ax}.
For metric tensor we
introduce
\be
ds^2_7 = e^{-6\phi} ds^2_4 + e^{4\phi} ds^2(\Sigma_3)\, ,
\label{a2}
\ee
where for definiteness we re-size $\Sigma_3$ here so that its Ricci tensor is $lg^2$ times
the metric tensor, i.e. with radius $\sqrt{2}/g$. Without losing generality,
we may rescale $l=1,0,-1$, each corresponding to $S^3,T^3,H^3$, respectively.
For the gauge field $SO(3)\subset SO(5)$,
we set
\be
A^{ab}_\1 = \tfrac{1}{g} \bar \omega^{ab}, \quad a,b=1,2,3 .
\ee
Here on the left hand side the $SO(3)$ indices refer to subalgebra $SO(3)\subset SO(5)_R$,
and on the right hand side the three-frame indices of $H^3$.  Furthermore, since the scalar fields
$T_{ij}$ should also respect our choice of $SO(5) \rightarrow SO(3)$,  we set
\be
T = {\rm diag} (e^{-4\lambda} , e^{-4\lambda}, e^{-4\lambda}, e^{6\lambda}, e^{6\lambda} )\,  .
\ee

Then it is straightforward to show that, when we substitute our ansatz above into the $D=7$ equations derived from the action \eqref{L7} we have a set of $D=4$ equations which
in turn can be derived by the following effective action.
\begin{align}
\frac{1}{\sqrt{g}} {\cal L}_4 & = R  - 30 (\partial\phi)^2 - 30 (\partial\lambda)^2
 - 3 g^2 \left[  e^{-10\phi} + \tfrac{1}{8} e^{8\lambda-14\phi} -\tfrac{1}{2} e^{-6\phi}
( e^{-8\lambda} + 4 e^{2\lambda} ) \right] .
\label{4L}
\end{align}
We also note that it is possible to employ a more general ansatz and obtain
the bosonic sector of ${\cal N}=2, \, D=4$ gauged supergravity with a vector multiplet
and two hypermultiplets. For more detail, readers are referred to \cite{Donos:2010ax}.

Considering the extremal points of the scalar potential above, one finds that there are
two distinct solutions. One is
\begin{align}
e^{-20\phi}&=2, \quad e^{10\lambda}=2 ,
\label{ss_sol}
\end{align}
which turns out supersymmetric. On the other hand,
the solution with
\begin{align}
e^{-20\phi} &= \frac{486}{625}, \quad e^{10\lambda}=10 \, ,
\end{align}
is not supersymmetric. In this paper we aim to identify the field theory dual of
the supersymmetric solution. It will be very interesting if we can also establish
the dual of the second solution.

Plugging \eqref{ss_sol} back into \eqref{4L}, we find that the curvature radius of the
supersymmetric $AdS_4$ solution is
\be
L^2 = \sqrt{2}/{g^2}  \, .
\ee
Now we make repeated  use of the uplifting formulae \eqref{a2} and \eqref{metel}
and obtain the following $D=11$ metric,
\begin{align}
ds^2_{11}
&
= \frac{2^{2/3} (1+\sin^2\theta)^{1/3}}{g^2}\Bigg[ ds^2(AdS_4) + ds^2(H^3)
+ \frac{1}{2}\left( d\theta^2
 +\frac{\sin^2\theta}{1+\sin^2\theta} d\phi^2\right)
\nn\\ &
 +\frac{\cos^2\theta}{1+\sin^2\theta}\sum_{a=1}^3
 (d\tilde \mu^a + \bar \omega^{a}_{\;\; b} \tilde\mu^b )^2
\Bigg] \, ,
\label{11metric}
\end{align}
where $\tilde \mu^a$ parametrize $S^2$, i.e. $\sum_{a=1}^3 (\tilde\mu^a)^2=1$.
In this expression we scaled both $AdS_4$ and $H^3$ to have unit radius, and
$0<\theta<\pi/2$. The
parameter $g$ can be related to the number of M5-branes $N$, through the
flux quantization condition. The four-form flux, restricted to the squashed four-sphere
$X_4$ parametrized by $\theta,\phi,\tilde\mu^a$, is
\be
G_\4 |_{X_4}= -\frac{8\pi^3}{g^3}d\left[
\frac{\cos^3\theta}{1+\sin^2\theta}
\right]
 \wedge d\phi \wedge {\rm vol} ( S^2) \, .
\label{gflux}
\ee
The M5-brane number is determined by integrating the above expression and using \eqref{Q5}.
\be
N = (\pi l_P^3 g^3)^{-1} .
\ee

The last step in the gravity computation is to use the general formula for holographic free energy for $AdS_4$ derived in \cite{Martelli:2011fu}.  For the $AdS_4$ dual of any ${\cal N}=2$ superconformal field theory
on $S^3_b$, the free energy is
\be
{\cal F}^{\rm gravity}_b = \frac{\pi}{8G_4}\left(b+\frac{1}{b}\right)^2 \, , \label{free enenergy for AdS(b)}
\ee
where $G_4$ is four-dimensional Gravitational constant.  $G_4$ is easily obtained from
the volume of the {\it internal} seven-dimensional space in \eqref{11metric}. The result is
\be
{\cal F}^{\rm gravity}_b = \frac{N^3}{12\pi}\left( b+\frac{1}{b} \right)^2 \textrm{vol}(H^3) \,  .
\ee

\section{Field theory analysis using 3d-3d correspondence} 
\label{sec : field theory}

\subsection{3d-3d relation and Chern-Simons theory}
The 3d-3d correspondence is a conjecture which relates the supersymmetric ptn for a 3d  $T_{N}[M]$ theory on a curved background $B$ to  topological  invariants of the manifold $M$. 
The topological invariants can be obtained by integrating out Kaluza-Klein modes of 6d $A_{N-1}$ (2,0) theory along $B$. 
When $B$ is a general squashed Lens space, 
\begin{align}
L(0):=S^2\times S^1 \textrm{ or }L(k):=S^3/\mathbb{Z}_k \textrm{ for } k \geq 1\;,
\end{align}
the corresponding topological invariants turns out to be  a $PGL(N)$ CS theory on the internal manifold \cite{Dimofte:2014zga}. 
The {\em complex} CS ptn is  defined by the following path-integral: 
\begin{align}
Z^{CS}_N(\hbar,\tilde{\hbar};M)=&\int [d\CA][d\overline{\CA}] \exp \left({\frac{1}{2\hbar}CS[\mathcal{A};M] +\frac{1}{2\tilde{\hbar}}}CS[\overline{\CA};M]\right)\;, \nn
\\
& \textrm{with }CS [\mathcal{A};M] := \int_M \left(\CA \wedge d\CA+ \frac{2}3 \CA^3 \right)\;. \label{CS path-integral}
\end{align}
The holomorphic  and anti-holomorphic  coupling constants are
\begin{align}
\frac{4\pi}{\hbar} = k+ \sigma;, \quad \frac{4\pi }{\tilde{\hbar}} = k-\sigma\;, 
\end{align}
where   $k$ is  integer and $\sigma$ is either real  or purely imaginary.   In the 3d-3d correspondence, the integer level $k$ is identified as the label $k$ of Lens space $L(k)$ and $\sigma$ is related to a squashing parameter. 
For example, consider the case when  the curved manifold is a squashed 3-sphere $S^3_b$ defined by
\begin{align}
S^3_b = \{ (z,w) \in \mathbb{C}^2 : b^2|z|^2 +\frac{1}{b^2 }|w|^2=1 \}\;. \label{squashed sphere}
\end{align}
The corresponding topological quantity is $PGL(N)$ CS ptn on $M$ with $k=1$ and $\sigma = \frac{1-b^2}{1+b^2}$ \cite{Cordova:2013cea} \footnote{In \cite{Cordova:2013cea}, $\sigma$  was given by $i \frac{1-b^2}{1+b^2}$. Here, following a recent work by Dimofte \cite{Dimofte:2014zga}, we ``erased the $i$".}.  
In terms of holomorphic and anti-holomorphic coupling, this corresponds to
\begin{align}
\hbar = 2\pi i (1+b^2)\;, \quad \tilde{\hbar} = 2\pi i (1+b^{-2})\;.
\end{align} 
This relation  looks  different from the original 3d-3d relation $\hbar =2\pi ib^2$ and $ \tilde{\hbar}:=2\pi ib^{-2}$. In the quantization of CS theory, however,  the more relevant parameters are exponentiated 
ones $q:=e^{\hbar}, \tilde{q}:=e^{\tilde{\hbar}}$ and the difference $2\pi i$ becomes irrelevant \cite{Dimofte:2014zga} . Thus, the 3d-3d relation says
\begin{align}
&\textrm{$Z_{S^3_b} (T_N[M])$, supersymmetric partition function of $T_{N}[M]$ on $S^3_b$ } \nn
\\
&=\textrm{$Z^{\textrm{CS}}_{N}(M;\hbar)$,  $PGL(N)$ CS ptn on $M$ with $\hbar =2 \pi i b^2$ and $\tilde{\hbar}=2\pi i b^{-2}$}\;.   \label{3d-3d dictionary 2}
\end{align}
We assume the following basic properties for the CS ptn $Z^{\textrm{CS}}_N$ whose correctness seems to be supported by several previous works \cite{Terashima:2011qi,Terashima:2011xe,Dimofte:2011ju,Andersen:2011bt}.
\begin{align}
1.\;&\textrm{In the limit $\hbar = 2\pi i b^2 \rightarrow 0$ with real $b$, $Z^{\textrm{CS}}_{N}(M;\hbar)$  has the same asymptotic expansion } \nn
\\
&\textrm{as the perturbative expansion of $PGL(N)$ CS ptn around a saddle point $\CA^{(\rm conj)}$.}
\label{Two things on CS ptn-1}
\\
2.\;&\textrm{It has the non-perturbative symmetry $\hbar \leftrightarrow -4\pi^2/\hbar$, }  Z^{\textrm{CS}}_N (\hbar)= Z^{\textrm{CS}}_N(-4\pi^2/\hbar).
\label{Two things on CS ptn-2}
\end{align}
The second property follows from the manifest symmetry $b\leftrightarrow b^{-1}$ of the squashed 3-sphere.   
The path-integral  is  not well-defined mathematically but its perturbative expansion around a given saddle point is well-defined.
For the CS theory, the saddle points are flat connections satisfying $d\CA +\CA\wedge \CA=0$.
Around a saddle point $\CA^{(\a)}$, the perturbative CS ptn can be expanded as
\begin{align}
Z^{(\a)}_{\textrm{pert}}(\hbar ;M) =\exp \left(\frac{1}{\hbar} S^{(\a)}_0  -\frac{3}{2} \log \hbar + S_1^{(\a)}+ \hbar S_2^{(\a)}+\cdots +\hbar^{n-1} S^{(\a)}_n+ \cdots\right)\;. \label{perturbative expansion for CS}
\end{align}
Each perturbative coefficient $\{S^{(\a)}_n\}$ is a topological invariant of $M$ and can be computed using standard field theoretic  techniques based on Feynmann diagrams.  For example,
\begin{align}
&S_0^{(\a)}= \frac{1}2 CS[\CA^{(\a)}]\;, \qquad \qquad \qquad \;  \textrm{(classical part)} \nn
\\
&S_1^{(\a)} = \frac{1}2 \log \textrm{Tor}_{\textrm{adj}}[M, \CA^{(\a)}]\;, \quad \quad \textrm{(one-loop)} \;. \label{classic, 1-loop CS invariants}
\end{align}
Here $\textrm{Tor}_{\rm R}[M, \CA^{(\a)}]$ is the  Ray-Singer torsion  of an associated vector bundle in a representation R $\in \textrm{Hom}\big{(}G\rightarrow GL(V_R)\big{)}$
twisted by a flat $G$-connection $\CA^{(\a)}$,
\begin{align}
\textrm{Tor}_{\rm R}[M, \CA^{(\a)}] =\frac{\big{[}\det'\Delta_0 ({\rm R}, \CA^{(\a)})\big{]}^{3/2}}{\big{[}\det'\Delta_1 ({\rm R}, \CA^{(\a)})\big{]}^{1/2}}\;.
\end{align}
Here $\Delta_n({\rm R},\CA^{(\a)})$ is a Laplacian acting on $V_{\rm R}$-valued $n$-form twisted by a  flat connection  $\CA^{(\a)}$.
Both of $S_0, S_1$ are mathematically well-defined geometrical quantities which have independent meaning.
Higher order invariants $S_n$ $(n\ge 2)$  have less obvious geometrical meaning but they can be rigorously defined using Feynman diagrams \cite{Dimofte:2009yn}.% \textcolor{red}{(Is the last statement obvious?)}

The squashed 3-sphere ptn for a 3d $\CN=2$ theory with a flavor symmetry of rank $r$ depends on $r$ complex parameters which combine real masses and R-charges.  For a knot complement $M=S^3\backslash \mathcal{K}$ with a full knot $\mathcal{K}$,  the $T_{N}[M]$ theory has $SU(N)$ flavor symmetry and the ptn depends on $(N-1)$ parameters.   The CS ptn function on knot complements $M$  also depends on $N-1$ parameters which parameterize boundary condition on $\partial M =\mathbb{T}^2$,
\begin{align}
\textrm{Hol}_{\textrm{meridian}} (\mathcal{A}) \sim  
\begin{pmatrix}
 e^{\mu_1} & 1 & 0 & 0 &  0 &\ldots  \\0 & e^{\mu_2}& 1 &0  & 0 & \ldots  \\0 & 0 & e^{\mu_3} & 1 & 0 &\ldots \\0 & 0 & 0 & e^{\mu_4} & 1 &\ldots  \\ \ldots & \ldots & \ldots & \ldots & \ldots &\ldots
\end{pmatrix}
 \,.
 \label{boundary condition}
\end{align}
Here, $\textrm{Hol}_{\gamma} (\mathcal{A})$ denotes a gauge holonomy  along a cycle $\gamma$ of gauge field $\mathcal{A}$.
We cannot impose Dirichlet boundary conditions on  holonomies  along two cycles simultaneously since they are canonically conjugate to each other.  To study  the AdS$_4$/CFT$_3$ for $T_{N}[M]$ theories, we  focus on the case when  the meridian holonomy is parabolic
\begin{align}
\mu_i=0 \quad (\textrm{parabolic  holonomy})
\end{align}
which corresponds to undeformed (with zero real mass) conformal $T_{N}[M]$ theory.
There are only finite number of flat connections $\{\CA^{(\a)}\}$  satisfying the boundary condition  \eqref{boundary condition} for given meridian variables $\{\mu_i\}$.
For $N=2$ and hyperbolic knot complement $M$, there exist two flat connections $\CA^{\textrm{(geom)}}_{N=2}$ and $\CA^{\textrm{(conj)}}_{N=2}$
which can be constructed from a unique complete hyperbolic metric on $M$ as follows,
\begin{align}
\CA^{\textrm{(geom)}}_{N=2} := \omega+ i e\;, \quad \CA^{\textrm{(conj)}}_{N=2} :=\overline{\CA^{\textrm{(geom)}}_{N=2}}= \omega- i e \;. \label{def of A(geom) and A(conj) N=2}
\end{align}
Here $\omega$ and $e$ are the spin-connection and the vielbein,  respectively. Both of them are locally $so(3)$ Lie algebra valued one-forms. 
For $N\geq 3$, the `geometrical' and 'conjugate' $PGL(N)$ flat connections are  
defined by embedding the corresponding connections at $N=2$ via the $N$-dimensional irreducible representation $\rho_N$ of $PGL(2)$.
\begin{align}
\CA^{\textrm{(geom)}}_{N} :=  \rho_N \left(\CA^{\textrm{(geom)}}_{N=2} \right)\;,\quad \CA^{\textrm{(conj)}}_{N} :=  \rho_N \left(\CA^{\textrm{(conj)}}_{N=2} \right) \;. \label{def of A(geom) and A(conj)}
\end{align}
The hyperbolic metric for a knot compliment  $S^3\backslash \CK$ around the knot can be written as $z^2 (dx^2+dy^2)+\frac{dz^2}{z^2}$ where the knot is located at $z=0$ and $x,y$ parametrize the longitude and meridian direction respectively. Using the metric, one can check check that meridian  holonomies for both of  $\CA_{N}^{\textrm{(geom)}}$ and $\CA_{N}^{\textrm{(conj)}}$ are parabolic at the boundary.  
One important characteristic of the geometrical flat connection (its conjugate flat connection) is that it has the  maximum (minimum) value of the imaginary part of the CS functional among all flat connections with parabolic meridian holonomy at the boundary.
\begin{align}
\textrm{Im}\left( CS[\CA^{(\textrm{conj})}_N]  \right)  \leq \textrm{Im}\left( CS[\CA^{(\a)}_N]  \right) \leq  \textrm{Im}\left( CS[\CA^{(\textrm{geom})}_N]  \right)\;.
\end{align}
The first (or second) equality only hold for $\a=\textrm{conj}$ (or $\textrm{geom}$). The maximum and minimum  values are
\begin{align}
\textrm{Im}\left( CS[\CA^{(\textrm{geom})}_N]  \right) = -\textrm{Im}\left( CS[\CA^{(\textrm{conj})}_N]  \right)  = \frac{1}3 N (N^2-1)\textrm{vol}(M)\;,
\label{classical action for geom,conj}
\end{align}
where $\textrm{vol}(M)$ is  the hyperbolic volume, the volume measured using the unique complete hyperbolic metric of the knot complement $M$. 
For $N=2$, it follows from a direct computation using eq.~\eqref{def of A(geom) and A(conj) N=2}. For general $N> 2$, it follows from a simple group theoretical fact that
\begin{align}
\textrm{Tr}[\rho_N (h_1) \rho_N (h_2)]= \frac{1}{6} N (N^2-1) \textrm{Tr}[h_1 h_2]\;,
\end{align}
where $h_1$ and $h_2$ are elements of the Lie algebra of $PGL(2)$.
%

%%%%%%%%%%%%%%%%%%%%%%%%%%%%%%%
\subsection{Conjecture on the large $N$ behavior of perturbative invariants} \label{subsection : conjecture}
%%%%%%%%%%%%%%%%%%%%%%%%%%%%%%%
Looking at the formula \eqref{gravity free energy} for the gravity free energy  carefully, we observe the following remarkable fact: the gravity free energy  has the same expansion in $b^2$ as the perturbative expansion of $CS$ ptn   \eqref{perturbative expansion for CS} under the identification $\hbar  =2\pi ib^2 $.
Combining the holographic principle, $\log |Z_{S^3_b}(T_N)|  = -\CF^{\rm gravity}_b$,  and the 3d-3d relation, $Z_{S^3_b}(T_N) = Z_N^{\textrm{CS}}(\hbar;M)$, we  obtain following predictions on the large $N$ behavior of  perturbative invariants $\{S^{(\textrm{conj})}_n\}.
$\footnote{It could be wrong due to an ``order of limits'' issue.  In AdS/CFT, we  take the limit $N\rightarrow \infty$ with fixed $\hbar$. But for CS ptn, we first asymptotically expand around $\hbar=0$ and then take the limit $N\rightarrow \infty$ on each expansion parameters $\{S_n\}$. We assume the uniform convergence of the large $N$ free energy and the issue does not matter. }
\paragraph{Conjecture} The $PGL(N)$ CS perturbative invariants $\{S^{(\textrm{conj})}_n(N)\}$ around the saddle point $\CA^{\textrm{(conj)}}_N$ on a hyperbolic 3-manifold $M$ have the following large $N$ behavior.
\begin{align}
&\lim_{N \rightarrow \infty }\frac{1}{N^3} \textrm{Im}[S^{(\textrm{conj})}_0] = - \frac{1}6 \textrm{vol}(M)\;, \nn
\\
&\lim_{N \rightarrow \infty }\frac{1}{N^3} \textrm{Re}[S^{(\textrm{conj})}_1] = - \frac{1}{6 \pi} \textrm{vol}(M)\;,  \nn
\\
&\lim_{N \rightarrow \infty }\frac{1}{N^3} \textrm{Im}[S^{(\textrm{conj})}_2] =  \frac{1}{24 \pi^2} \textrm{vol}(M)\;, \nn
\\
&\lim_{N \rightarrow \infty }\frac{1}{N^3} \textrm{Re}[S^{(\textrm{conj})}_{2j-1}] =\lim_{N \rightarrow \infty }\frac{1}{N^3} \textrm{Im}[S^{(\textrm{conj})}_{2j}]=  0\;,\quad j=2,3,\ldots, \infty \;. \label{conjecture}
\end{align}
We use the gravity free energy calculation in eq.~\eqref{gravity free energy} and one of our assumption on  $Z^{\textrm{CS}}_{N}$ given in eq.~\eqref{Two things on CS ptn-1}. For the classical part $S^{(\textrm{conj})}_0$, its behavior  can be easily understood from eq.~\eqref{classic, 1-loop CS invariants} and \eqref{classical action for geom,conj}. For the one-loop part $S^{(\textrm{conj})}_1$, its large $N$ behavior can be derived using a mathematical theorem found  in  \cite{2011arXiv1110.3718M}.
\begin{align}
\lim_{m\rightarrow \infty}\frac{1}{m^2} \log \textrm{Tor}_{\rho_m}[M, \mathcal{A}^{\textrm{(geom)}}_{N=2}]= - \frac{1}{4\pi} \textrm{vol}(M)\;,
\end{align}
where $\rho_m$ is the irreducible $m$-dimensional representation of $PGL(2)$. Applying the theorem to $S^{\textrm{(conj)}}_1= \overline{S^{\textrm{(geom)}}_1}$ in \eqref{classic, 1-loop CS invariants} using the branching rule $\textbf{adj}= \rho_{3}\oplus \rho_{5} \oplus \ldots \oplus \rho_{2N-1}$, we arrive at
\begin{align}
S_1^{(\textrm{conj})}(N) &=  - \frac{1}{4\pi } \textrm{vol}(M) \left(3^2+5^2+\ldots + (2N-1)^2 \right)+(\textrm{subleading in }1/N) \nn
\\
&=  -\frac{1}{6\pi }\textrm{vol}(M) N^3 +(\textrm{subleading})\;, \nn
\end{align}
which is compatible with the conjecture.  We currently have little analytic understanding of the loop invariants $S_n$ $(n \geq 2)$.
\paragraph{Relation to Volume Conjecture} The original volume conjecture (VC)  \cite{1996q.alg.....1025K,1999math......5075M} relates an asymptotic limit of colored Jones polynomial $J_R (q;\CK)$ of a knot $\mathcal{K}$ to the hyperbolic volume of  knot complement  $S^3\backslash \mathcal{K}$.
\begin{align}
\lim_{n=k\rightarrow \infty} \frac{2\pi }k \log \left|\frac{J_n (q=e^{2\pi i/k};\CK)}{J_n(q=e^{2\pi i/k};\bigcirc)} \right| = \textrm{vol}(S^3\backslash \CK)\;.
\end{align}
Here $\bigcirc$ denotes an unknot. Refer to \cite{Dimofte:2010ep} for review on VC and its generalizations.  The colored Jones polynomial is colored by a $SU(2)$ representation $R$ and $J_n (q;\CK):= J_{R=\rho_n}(q;\CK)$. The original definition of the polynomial is given in an algebraic and  combinatorial way.  Witten gave the following alternative definition based on a path-integral \cite{Witten:1988hf}.
\begin{align}
J_R (q;\CK) = \frac{\int [dA]  \exp  \left(i k_{\textrm{bare}}CS[A;S^3] \right) \textrm{Tr}_R \left(\textrm{Hol}_{\CK}(A) \right) } {\int [dA]   \exp \left(i k_{\textrm{bare}}CS[A;S^3] \right)  }\;, \quad q=e^{2\pi i/k} (k:=k_{\textrm{bare}}+2)\;. \nn
\end{align}
Here $A$ is an $SU(2)$ gauge field on $S^3$. Using the path-integral definition of the colored Jones polynomial, a heuristic physical understanding of the VC can be given \cite{Gukov:2003na} and VC can be generalized to include sub-leading terms in the asymptotic limit \cite{Gukov:2006ze}. The generalized VC is
\begin{align}
 \log \frac{J_n (q;\CK)}{J_n(q;\bigcirc)}\sim  \frac{1}{\hbar} S_0^{(\textrm{geom})}(N=2)- \frac{3}2 \log \hbar+S_1^{(\textrm{geom})} (N=2)+\ldots\;,
\end{align}
in the limit  $\hbar :=\log q  \rightarrow 0$ and $n\rightarrow \infty$ with fixed $n\hbar  = 2\pi i$.

Jones polynomial can also be generalized by replacing the gauge group $SU(2)$ by $SU(N)$,
\begin{align}
J^N_R (q;\CK) = \frac{\int [dA]  \exp  \left(i k_{\textrm{bare}}CS[A;S^3] \right) \textrm{Tr}_R \left(\textrm{Hol}_{\CK}(A) \right) } {\int [dA]   \exp \left(i k_{\textrm{bare}}CS[A;S^3] \right)  }\;, \quad q=e^{2\pi i/k} (k:=k_{\textrm{bare}}+N)\;, \nn
\end{align}
where $A$ is a $SU(N)$ gauge field on $S^3$ and $R$ is a representation of $SU(N)$. Considering an asymptotic limit of the $SU(N)$ quantum invariant, the VC can be further generalized as follows \cite{Dimofte:2010ep}
\begin{align}
 \log \frac{J^N_R (q;\CK)}{J^N_R(q;\bigcirc)}\sim  \frac{1}{\hbar} S_0^{(\textrm{geom})}(N)- \frac{3}{2}\log \hbar+S_1^{(\textrm{geom})} (N)+\ldots\;,
\label{VC-SU(N)}
\end{align}
in an asymptotic limit  $\hbar :=\log q \rightarrow 0$ and $(\delta^*+\lambda^*_R)\rightarrow  \textrm{diag}(\infty,\infty,\ldots, \infty)$ with fixed $\exp \left(\hbar (\delta^*+\lambda^*_R)\right)= \mathbb{I}_N$. $\delta$ is half of sums of positive roots of $SU(N)$ and $\delta^*$ is its dual element.  The dual element is defined using the nondegenerate trace form $-\textrm{Tr}$ as an inner product. $\lambda_R$ is the highest weight vector of a representation $R$. $\delta^*$ and $\lambda_R^*$ are elements of the  Cartan subalgebra of $SU(N)$.

The formula \eqref{VC-SU(N)} says that  the asymptotic  expansion of the  $SU(N)$ invariant is determined by the perturbative $PGL(N)$ CS invariants $\{ S^{(\textrm{geom})}_n \}$. Then, what determine the asymptotic growth rate of these perturbative CS invariants as $N$ goes to infinity? Our conjecture \eqref{conjecture} gives the answer to the question. Using the simple fact $S^{(\textrm{geom})}_n = \overline{S^{(\textrm{conj})}_n}$,
\begin{align}
&\lim_{N \rightarrow \infty }\frac{1}{N^3} \textrm{Im}[S^{(\textrm{geom})}_0] =  \frac{1}6 \textrm{vol}(M)\;, \nn
\\
&\lim_{N \rightarrow \infty }\frac{1}{N^3} \textrm{Re}[S^{(\textrm{geom})}_1] = - \frac{1}{6 \pi} \textrm{vol}(M)\;,  \nn
\\
&\lim_{N \rightarrow \infty }\frac{1}{N^3} \textrm{Im}[S^{(\textrm{geom})}_2] =  -\frac{1}{24 \pi^2} \textrm{vol}(M)\;, \nn
\\
&\lim_{N \rightarrow \infty }\frac{1}{N^3} \textrm{Re}[S^{(\textrm{geom})}_{2j-1}] =\lim_{N \rightarrow \infty }\frac{1}{N^3} \textrm{Im}[S^{(\textrm{geom})}_{2j}]=  0\;,\quad j=2,3,\ldots, \infty \;. \label{conjecture2}
\end{align}
According to the conjecture, the leading large $N$ behavior of the perturbative invariants is determined by the hyperbolic volume.

\subsection{Free energy at finite $b$}
So far we have analyzed the free energy, 
\[
\CF(\hbar;M)=-\log |Z_{S^3_b}(T_{N}[M])|
=-\log |Z^{\textrm{CS}}(\hbar;M)|\,,
\] 
in an asymptotic limit  $b\rightarrow 0$.  What can we say about the free energy at  finite $b$?  Consider the leading $N^3$-coefficient  of the free energy
 \begin{align}
 \CF_{N^3}(\hbar;M):= -\lim_{N\rightarrow \infty} \frac{\log |Z^{\textrm{CS}}(\hbar;M)|}{N^3}\;.
 \end{align}
 If our conjecture \eqref{conjecture} is true, we have the following  asymptotic expansion
 \begin{align}
\CF_{N^3}(\hbar;M)\rightarrow  \frac{1}{12\pi}(b^{-2}+2+b^2+0 b^4+\ldots+0 b^{2n}+\ldots) \textrm{vol}(M)\;,  \label{asymptotic for F_{N^3}}
 \end{align}
when $\hbar=2\pi i b^2 \rightarrow 0$ goes to zero with real $b$. Note that the symmetry $\hbar\leftrightarrow -4\pi^2/\hbar$ (or equivalently $b\leftrightarrow b^{-1}$)  is manifest  in the perturbative expansions of $\CF_{N^3}$.
Although the symmetry is an expected property of $\CF_{N^3}$ from \eqref{Two things on CS ptn-2},
its appearance in the perturbative expansions is somewhat unexpected.
The symmetry is a non-perturbative property of the free energy and we expect that it can be seen only after taking into account of
all non-perturbative corrections of the form $e^{- \frac{4\pi ^2}{\hbar}}$ as well as all perturbative corrections.
Actually the symmetry of the full free energy $\CF=-\log |Z^{\textrm{CS}}|$ is not seen in the perturbative expansions
but only can be  seen after taking into account of   non-perturbative corrections together.
The existence of the symmetry in the perturbative expansions of  $\CF_{ N^3}$ strongly suggest that
there is no non-perturbative corrections to $\CF_{N^3}$, or equivalently non-perturbative corrections of $\CF$ are suppressed by $1/N$,
and the perturbative expansion \eqref{asymptotic for F_{N^3}} is actually a convergent series. Thus,
\begin{align}
\CF(\hbar;M)= \frac{N^3}{12\pi } (b+b^{-1})^2 \textrm{vol}(M)+(\textrm{subleading in $1/N$})\;, \label{field theory free energy}
\end{align}
which  perfectly matches the gravity calculation \eqref{gravity free energy}.  In the derivation of the above  result,  we seriously use  our  assumptions on $Z^{\textrm{CS}}(\hbar;M)$ \eqref{Two things on CS ptn-1}, \eqref{Two things on CS ptn-2} and our conjecture \eqref{conjecture}. The conjecture for $S_0$ and $S_1$ were proven in section \ref{subsection : conjecture}.   In the next section, we will  numerically confirm the conjecture on $S_2, S_3$ for various knot complements  using Dimofte's state-integral model.

%%%%%%%%%%%%%%%%%%%%%%%%%%%%
\section{State-integral model for $PGL(N)$ CS ptn} 
\label{sec : dimofte-state integral}

In this section, we calculate the perturbative CS invariants $\{ S^{\textrm{(\textrm{conj})}}_n \}$ for various knot complements $M$ using Dimote's state-integral model  \cite{Dimofte:2011gm,Dimofte:2012qj,2011arXiv1111.2828G,2012arXiv1207.6711G,Dimofte:2013iv,Garoufalidis:2013upa}. 
From the calculation we numerically check the conjectures \eqref{conjecture} up to $n=3$. The state-integral model gives finite dimensional integral expression for the  $PGL(N)$ CS ptn $Z^{\textrm CS}_N[\hbar;M]$  using an ideal triangulation $\CT$ of $M$.   It was verified in \cite{Dimofte:2012qj} that the perturbative expansion of the state-integral model  reproduces the known perturbative invariants $S_0$  and $S_1$ for various knot complements $M$.
In fact, the integral should be interpreted as a {\em contour integral}. Depending on the choice of contour it will give different topological invariants of $M$. The state-integral model for $M$ contains a product of quantum dilogarithm functions and Gaussian factors.
Formally it looks exactly the same as the integral expression for an $S^3_b$-ptn for a 3d $\CN=2$ abelian CS-matter theory obtained by a localization method \cite{Kapustin:2009kz,Hama:2011ea}.  The 3d theory is identified as  the $T_{N}[M]$ theory in \cite{Dimofte:2011ju}. In the integral from localization, the contour runs along the real axis since the saddle points are constant modes of  real scalars $\sigma$ in $\CN=2$ vector multiplets.  Thus, the state-integral will give the topological invariant $Z^{\textrm{CS}}(\hbar;M)$ in the 3d-3d correspondence \eqref{3d-3d dictionary 2} if the contour runs along the real axis.\footnote{More precisely, the contour should lie slightly above the real axis in order not to touch a singularity at  the origin, $X=0$.} Using the contour prescription for the state-integral we can test the our assumption \eqref{Two things on CS ptn-1}.  For the self-completeness, we start with reviewing  basic ideas  of the state-integral model.

\subsection{Dimofte's state integral model}
An important ingredient in the study of a CS theory on a 3-manifold $M$ with boundary is a phase space $\CP(\partial M)$ canonically associated to the boundary $\partial M$ of $M$ \cite{Dimofte:2009yn}.  For $PGL(N)$ CS theory, the phase space is mathematically described as
\begin{align}
\CP_N (\partial M) =  \{\textrm{space of flat $PGL(N)$ connections on $\partial M$}\}/(\textrm{gauge equivalence})\;.
\end{align}
The boundary phase space has  a natural symplectic structure,
\begin{align}
\omega_{\partial M} = \frac{1}{\hbar}\int_{\partial M}\textrm{Tr}(\delta \CA \wedge \delta \CA)\;,
\end{align}
where $\delta \CA$ is an infinitesimal variation of the $PGL(N)$ gauge field $\CA$ restricted on $\CP(\partial M)$.
Those flat $PGL(N)$ connections on $\partial M$ which can be extended over the whole $M$ as flat connection form 
a Lagrangian submanifold of $\CP(\partial M)$, denoted by $\mathcal{L}_N (M) \subset \CP_N (\partial M)$:
\begin{align}
\CL_N (M)= \{\textrm{space of flat $PGL(N)$ connections on $M$}\}/(\textrm{gauge equivalence})\;.
\end{align}
A related  problem is to find the Lagrangian $\mathcal{L}_N (M) \subset \CP_N (\partial M)$ and to quantize it for general 3-manifold $M$. 
For $PGL(2)$, the problem was systematically  studied in \cite{Dimofte:2011gm} using an ideal triangulation  of $M$,
\begin{align}
M= \left(\bigcup_{i=1}^{k} \Delta_i \right)/\mbox{(gluing data)} \;. \label{tetrahedral decom of M}
\end{align}
The gluing data dictates which edges from which tetrahedra ($\Delta$) should be identified.
In \cite{Dimofte:2011gm}, it was demonstrated that  $P_2(\partial M)$, $\CL_2 (M)$ and its quantization $\hat{\CL}_2(M)$ can be constructed by `gluing' the datum of each tetrahedron $\Delta_i$. For example, the boundary phase space $\CP_2(\partial M)$ can be constructed by the symplectic reduction,
\begin{align}
\CP_2 (\partial M) = \prod_{i=1}^k \CP_2(\partial \Delta_i) //\{C_I = 0 \}\;. \label{boundary phase from a reduction}
\end{align}
For a single tetrahedron $\Delta_i$, we assign the following `elementary' phase space,
\begin{align}
&\CP_2 (\partial \Delta_i) = \{ (Z_, Z'_i, Z''_i) : Z_i +Z'_i+Z''_i = i \pi \}\quad \textrm{with symplectic structure} \nn
\\
&\{Z,Z'\}=\{Z',Z''\}=\{Z'',Z\}=\hbar\;.  \label{tetrahedron phase space for N=2}
\end{align}

\begin{figure}[h!]
\begin{center}
   \includegraphics[width=0.25\textwidth]{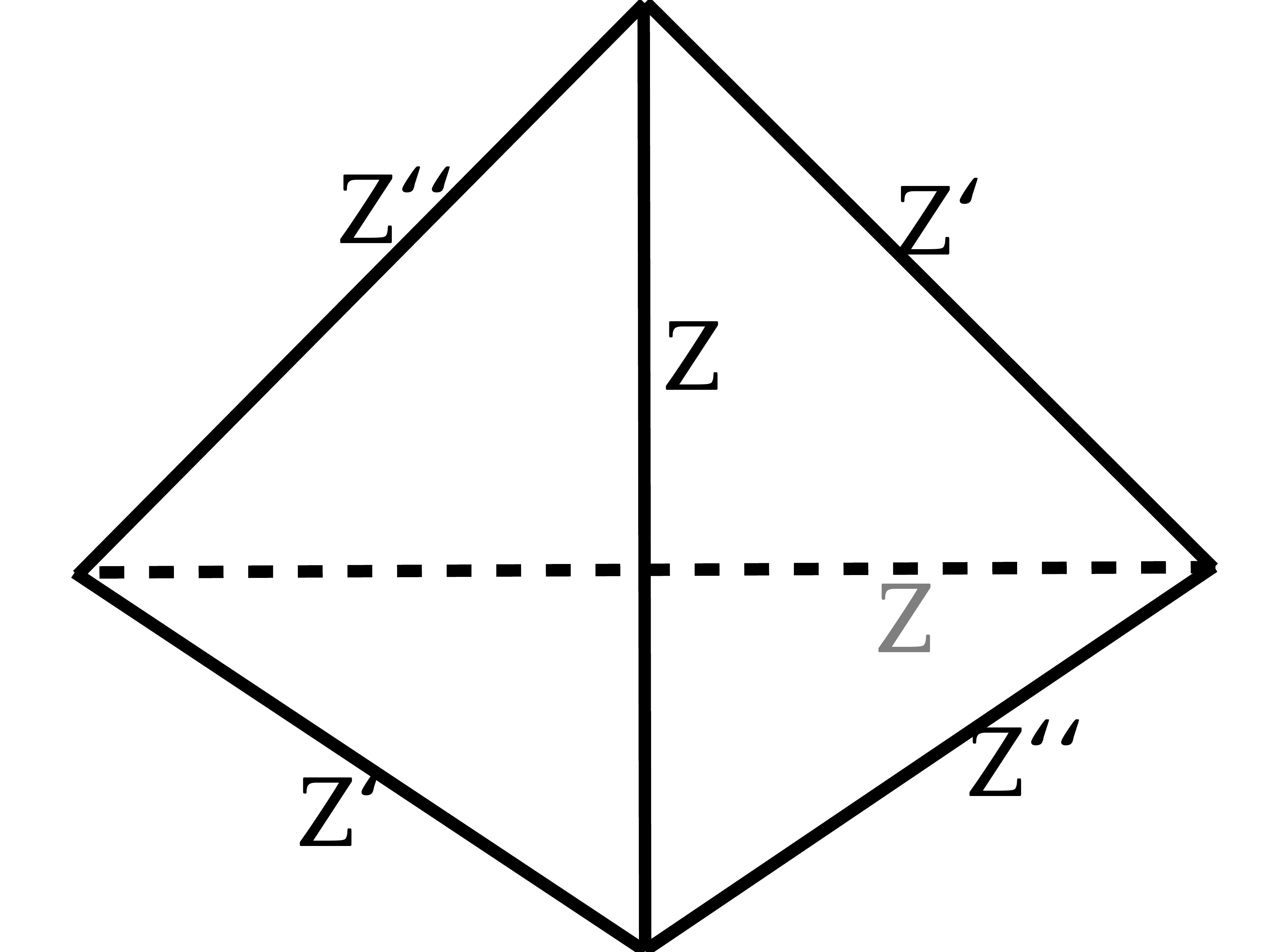}
   \end{center}
   \caption{Edge variables $(Z,Z',Z'')$ assigned to pairs of edges in a tetrahedron. }
    \label{fig:edge-parameters}
\end{figure}

\noindent
The variables $(Z,Z',Z'')$ assigned to each pair of edges in $\Delta$.  Upon the identification \eqref{def of A(geom) and A(conj) N=2}, 
the imaginary parts of the edge variables  geometrically represent the dihedral angles between the two faces meeting at each edge. 
The internal edges $\{ C_I \}$ can be expressed as linear sums of these edge variables according to the gluing data.
\begin{align}
C_I = \textrm{(sum over all variables for edges meeting at the $I$-th internal edge)} - 2\pi i\;.
\end{align}
The condition $C_I=0$ guarantees that there is no conical singularity around each internal edge. 
Classically, the Lagrangian $\CL_2 (M)$ is the image of the product of the elementary Lagrangians 
\begin{align}
\CL_2 (\Delta_i) = \{ e^{Z_i''}+e^{-Z_i}-1=0 \}
\end{align}
under the symplectic reduction \eqref{boundary phase from a reduction}.
Upon quantization, the edge variables $(Z_i, Z'_i, Z''_i)$ are promoted to operators $(\hat{Z}_i, \hat{Z}'_i, \hat{Z}''_i)$ acting on a suitable Hilbert space $\CH (\partial \Delta_i)$. The quantum counterpart of the symplectic reduction procedure \eqref{boundary phase from a reduction} produces the quantum Lagrangian $\hat{\CL}_2(M)$ from its building blocks 
\begin{align}
\hat{\CL}_2(\Delta_i)  = \{ e^{\hat{Z}''_i} +e^{-\hat{Z}_i}-1 \simeq 0 \}\;. \label{quantum lagrangian for tetrahedron} 
\end{align}
An operator equation `$\hat{A}\simeq 0$' means that there exists some state (ket-vector) $|\Psi(M) \rangle  \in \mathcal{H}(\partial M)$ such that $\hat{A} | \Psi (M)\rangle =0$ holds.

Instead of reviewing the  quantum symplectic reduction for $\hat{\CL}_2(M)$, we will review the reduction for the wave-function $|\Psi_{N=2} (M)\rangle$ annihilated by the operators contained in $\hat{\CL}_2(M)$. In a polarization $\Pi =(\chi_\a,  \xi_\a )|_{\a=1, \ldots ,\half \textrm{dim} \CP (\partial M)}$ of $\CP(\partial M)$, a choice of decomposition of the coordinates on the phase space into ``canonical coordinates''  $\chi$ and ``canonical momenta'' $\xi$, we define the $PGL(2)$ ``CS wave-function" as
\begin{align}
Z^{\rm CS}_{N=2} (M; \Pi) (\chi_\a) = \langle  \chi_\a ;\Pi  |\Psi_{N=2}(M)\rangle \;.
\end{align}
Here $| \chi_\a ;\Pi\rangle$ denote a position basis for the Hilbert-space $\CH_{N=2} (\partial M)$ in the polarization $\Pi=(\chi,\xi)$. 
In the path-integral representation of the CS wave-function \eqref{CS path-integral}, 
a choice of polarization determines which components of gauge fields should be fixed at the boundary.  
For a single tetrahedron, the CS wave-function in the polarization $\Pi_{Z} := (Z,Z'')$ is given by a quantum dilogarithm function (QDL) 
\cite{Dimofte:2011gm},
\begin{align}
&Z^{\rm CS}_{N=2} (\Delta;\Pi_Z) (Z) = \psi_{\hbar} (Z)\,,\; \textrm{where} \nn
\\
&\psi_{\hbar} (Z)=  \left(\begin{array}{c}  \prod_{r=1}^\infty \frac{ 1- q^{r} e^{-Z}}{1- \tilde{q}^{-r+1}e^{-\tilde{Z}}} \quad |q|<1\\ \prod_{r=1}^\infty \frac {1- \tilde{q}^{r}e^{-\tilde{Z}}}{ 1- q^{-r+1} e^{-Z}} \quad \quad |q|>1
\end{array}\right. \label{Q.D.L}
\end{align}
with
\begin{align}
q:= e^\hbar\;, \quad \tilde{q}:=e^{\tilde{\hbar}}:=e^{- \frac{4\pi^2}\hbar}\;, \quad \tilde{Z}=\frac{2\pi i}{\hbar}Z\;.
\end{align}
The QDL enjoys several interesting properties. It has following S-duality 
\begin{align}
\psi_{\hbar}(Z) = \psi_{\tilde{\hbar}}(\tilde{Z}) \;. \label{S-duality for QDL}
\end{align}
It satisfies following difference equations
\begin{align}
&(e^{\hbar \partial_{Z}}+e^{-Z} -1) \psi_{\hbar}(Z) =0\;, \nn
\\
&(e^{\tilde{\hbar} \partial_{\tilde{Z}}}+e^{-\tilde{Z}} -1) \psi_{\hbar}(Z) =0\;. \label{Differences for QDL}
\end{align}
This property reflects the fact that the wave-function $|\Psi_{N=2}(\Delta)\rangle$ is annihilated by the quantum Lagrangian in \eqref{quantum lagrangian for tetrahedron}.
The function is a meromorphic function with infinitely many simple poles and simple zeros located at
\begin{align}
&\textrm{simple zeros : } 2\pi i \mathbb{Z}_{> 0} +\hbar \mathbb{Z}_{> 0} \;, \nn
\\
&\textrm{simple poles : }  2\pi i \mathbb{Z}_{\leq 0} +\hbar \mathbb{Z}_{\leq 0} \;. \label{poles and zeros of QDL}
\end{align}
For other interesting properties of the QDL, see, {\it e.g.}, section 3.3 in \cite{Dimofte:2009yn}.
Polarizations $\Pi_{Z'}:=(Z', Z)$ and $\Pi_{Z''}:=(Z'',Z')$ are equvalent to $\Pi_{Z}$ up to cyclic re-labelling of edge variables. Therefore,
\begin{align}
Z^{\rm CS}_{N=2} (\Delta;\Pi_{Z'}) (X) = Z^{\rm CS}_{N=2} (\Delta;\Pi_{Z''}) (X) =Z^{\rm CS}_{N=2} (\Delta;\Pi_Z) (X) \;.
\end{align}
To obtain the CS wave-function $Z^{\rm CS}_{N=2}(M;\Pi)(\chi_\alpha)$ using the  tetrahedral decomposition \eqref{tetrahedral decom of M},
we first prepare the product of QDL's for all $k$ tetrahedra,
\begin{align}
Z^{\rm CS}_{N=2} \left(\bigcup_{i=1}^k \Delta_i;\Pi_{{\bf X}_\Delta}\right)(X_i)=\prod_{i=1}^k \psi_{\hbar } (X_i)\;.  \label{step 1 for PGL(2) CS}
\end{align}
Here $\Pi_{{\bf X}_\Delta}$ denote the collection of the polarisation choice for the boundary phase space of each tetrahedron $\Delta_i$, where the position variable $({\bf X}_{\Delta})_i$ is one of $Z,Z'$ and $Z''$ and the momentum $({\bf P}_{\Delta})_i$ is $Z'',Z$ and $Z'$, respectively.
As a second step, we perform a polarization transformation from $\Pi_{{\bf X}_\Delta}$ to a new one $\Pi_{\widetilde{\bf X}}=(\widetilde{\bf X},\widetilde{\bf P})$ where
\begin{align}
&\widetilde{\bf X}  = \{ C_I , \chi_\a\}\;, \quad \widetilde{\bf P} = \{ \Gamma_I, \xi_\a\} \;. \end{align}
Here $\{C_I\}$ are internal edge variables and  $\{\Gamma_I\}$ are its conjugate momenta. Recall that  $\{\chi_\a \}$ and $\{\xi_\a\}$ are position and momentum variables in our desired polarization $\Pi$. The index $I$ runs  $1,\ldots , k-\half \textrm{dim}\CP(\partial M)$ and $ \a=1,\ldots, \textrm{dim}\CP(\partial M)$. In general, there are more than $k-\half \textrm{dim}\CP(\partial M)$ internal edges but only that number of internal edges are linearly independent. The polarization change can be accomplished by a combination of  $Sp(2k,\mathbb{Z})$ transformation and affine shifts,
\begin{align}
\left(\begin{array}{c} \widetilde{\bf X}\\\hline  \widetilde{ \bf P} \end{array}\right):=
\left(\begin{array}{c} \bf{C}   \\ \vec{\chi}   \\\hline \bf{\G} \\  \vec{\xi} \end{array}\right) = \left(\begin{array}{c|c}A & B \\\hline C &  D\end{array}\right) \left(\begin{array}{c} {\bf X}_{\Delta} \\\hline {\bf P}_{\Delta} \end{array}\right)  - i\pi \left(\begin{array}{c} \nu \\\hline  \nu_p \end{array}\right) \label{Sp(2k,Z)+affine shifts}\;.
\end{align}
Here $\bf{C}$ denotes a vector of size $k-\half \textrm{dim}\CP(\partial M)$ whose elements are $C_I$, and $\vec{\chi}$ a vector of size $\half \textrm{dim}\CP(\partial M)$ whose elements are $\chi_\a$. Other vectors in the expression are defined in a similar way. The four $k\times k$ matrices $A,B,C,D$ form a $Sp(2k,\mathbb{Z})$ matrix. 
Under the polarization transformation, the CS wave-function transforms as  (see appendix C in \cite{Dimofte:2012qj} for the derivation)
\begin{align}
&Z^{\rm CS}_{N=2}  \left(\bigcup_{i=1}^k \Delta_i; \Pi_{\widetilde{\bf X}}\right)(\widetilde{\bf X}) =  \frac{1}{\sqrt{\det B}} \int \frac{d^k {\bf X}}{(2\pi i \hbar)^{k/2}} \exp \left[ - \frac{1}\hbar \widetilde{\bf X} \cdot (i \pi + \frac{\hbar}2) \nu_p \right. \nn
\\
& \qquad +\frac{1}{2\hbar} \bigg( (\widetilde{\bf X}+(i \pi +\frac{\hbar}2) \nu)\cdot D B^{-1} (\widetilde{\bf X}+(i \pi + \frac{\hbar}2\nu)) - 2 {\bf X}\cdot B^{-1} (\widetilde{\bf X}+(i \pi +\frac{\hbar}2)\nu ) \nn
\\
& \qquad \qquad \qquad +{\bf X}\cdot  B^{-1} A  {\bf  X} \bigg) \bigg] Z^{\rm CS}_{N=2}  \left(\bigcup_{i=1}^k \Delta_i;\Pi_{{\bf X}_\Delta} \right) ({\bf X})\;. \label{step 2 for PGL(2) CS}
\end{align}
Finally, applying the symplectic reduction $\{C_I=0\}$ to the wave-function, we obtain the CS wave-function for $M$,
\begin{align}
Z^{\rm CS}_{N=2}  (M; \Pi)(\vec{\chi}) = Z^{\rm CS}_{N=2}  \left(\bigcup_{i=1}^k \Delta_i;\Pi_{\widetilde{\bf X}}\right) \left({\bf C}={\bf 0}; \vec{\chi}\right)\;. \label{step 3 for PGL(2) CS}
\end{align}
This formula is subject to some ambiguities, such as the arbitrariness for the choice of ${\bf X}_{\Delta}$ and the choice of ${\bf C}$ appearing in $\widetilde{\bf X}$.
As a consequence, the CS wave-function is defined up to an overall pre-factor of the following form (see appendix C.5 in \cite{Dimofte:2012qj})
\begin{align}
\exp\left(\frac{\pi^2}{6\hbar} l + \frac{i\pi}4 m +\frac{\hbar}{24} n\right)\;, \quad l,m,n \in \mathbb{Z}\;. \label{ambiguity in state-integral}
\end{align}
Note that this factor is a pure phase  when $\hbar = 2\pi i b^2 $  with real $b$. Since we will compare the absolute value of $Z^{\rm CS}$ with the gravity free energy, this ambiguity can be ignored.

\begin{figure}[h!]
\begin{center}
   \includegraphics[width=.9\textwidth]{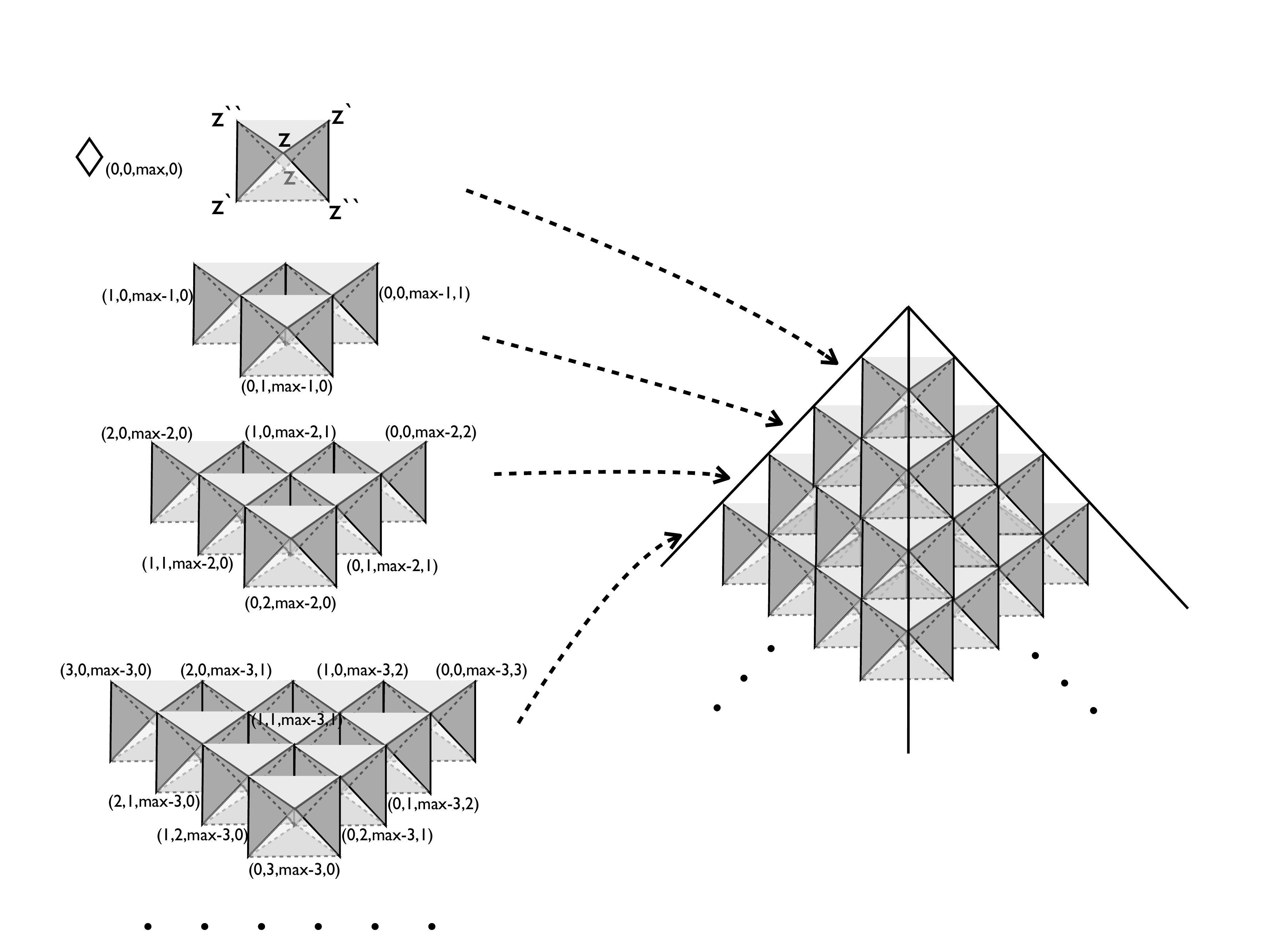}
   \end{center}
   \caption{$\frac{1}6 N(N^2-1) = 1+(1+2)+(1+2+3)+\ldots +(1+\ldots +N-1)$ octahedra form a single $PGL(N)$-tetrahedron. The octahedra $\Diamond_{(a,b,c,d)}$ are labelled by four non-negative integers $(a,b,c,d)$ with $a+b+c+d=\textrm{max}:=N-2$. The labeling rule can be understood from the figure.}
    \label{fig:N-decomposition}
\end{figure}

For general $PGL(N)$, we can still use the tetrahedral decomposition \eqref{tetrahedral decom of M}. But, $\CP_N (\partial \Delta)$ and $\CL_N (\Delta)$ are not so simple as in $\CP_2 (\partial \Delta)$ and $\CL_2 (\Delta)$ \eqref{tetrahedron phase space for N=2}. To construct $\CP_N (\partial \Delta)$ and $\CL_N (\Delta)$, we decompose a single tetrahedron $\Delta$ into a pyramid of  $\frac{1}6 N(N^2-1)$ octahedra $\Diamond$, as illustrated in Figure \ref{fig:N-decomposition}.
This is a 3d uplift of Fock and Goncharov's construction \cite{2003math.....11149F} on a Riemann surface  and  called $N$-decomposition (or octahedral decomposition) in \cite{Dimofte:2013iv}. Fock-Goncharov (FG) coordinates parameterize $PGL(N)$ flat connection moduli space on a Riemann surface using a tessellation for each triangle in a  triangulation of the surface.  Quantization of the moduli space with a suitable symplectic form defines a cluster algebra where the FG coordinates serve as so-called $y$-variables.  An ideal tetrahedron is generated by
acting a  flip in a 2d triangulation. In terms of the cluster algebra, a single flip  corresponds to sequence of $\frac{1}6 N (N^2-1)$ mutations and each mutation corresponds to octahedra in the $N$-decomposition.
A single octahedron has the same basic (quantum) geometrical data as a single  $PGL(2)$ tetrahedron.
\begin{align}
\CP(\partial \Diamond) = \CP_{N=2}(\partial \Delta)\;,
\quad  \CL( \Diamond) = \CL_{N=2}( \Delta)\;,
\quad \hat{\CL}(\Diamond) = \hat{\CL}_{N=2}( \Delta)\;.
\end{align}
A difference is that the boundary phase space coordinates  $\{Z,Z',Z''\}$ are associated to pairs of vertices of an octahedron as shown in Figure \ref{fig:N-decomposition}.
The $\frac{1}6 N(N^2-1)$ octahedra  in a $PGL(N)$ tetrahedron are labelled by four non-negative integers $(a,b,c,d)$ whose total sum is $N-2$.
The labeling rule can be understood from Figure \ref{fig:N-decomposition}.
We can also read off the `internal vertices' $\{ C_{a,b,c,d} \}_{a-1,b,c,d-1 \geq 0}^{ a+b+c+d=N-2}$, where the vertices of several octahedra meet:
\begin{align}
C_{a,b,c,d} =& Z_{a, b, c, d} + Z_{a - 1, b + 1, c + 1, d - 1} +
 Z'_{a - 1, b, c + 1, d}  \nn
 \\
 &+ Z'_{a, b + 1, c, d - 1} +
 Z''_{a, b, c + 1, d - 1}+ Z''_{a - 1, b + 1, c, d} - 2\pi i\;.
\end{align}
In total, there are $\frac{1}6 (N-3)(N-2)(N-1)$ linearly independent internal vertices.
Using the $N$-decomposition, the $PGL(N)$ boundary phase space $\CP_N (\partial \Delta)$  can be constructed as
\begin{align}
\CP_N (\partial \Delta) = \left( \prod_{(a,b,c,d)}\CP(\partial \Diamond_{(a,b,c,d)}) \right) //\{C_{a,b,c,d} = 0 \}\;.
\end{align}
The dimension of the phase space is $2\times (\frac{1}6 N(N^2-1)- \frac{1}6 (N-3)(N-2)(N-1) )=2(N-1)^2$.
For a 3-manifold $M$ composed of $k$ tetrahedra, we need $\frac{k}{6} N (N^2-1)$ octahedra to construct $\CP_N(\partial M )$.
\begin{align}
\CP_N (\partial M) = \left( \prod_{i=1}^k \prod_{(a,b,c,d)}\CP(\partial \Diamond^{(i)}_{(a,b,c,d)}) \right) //\{C_{*} =0\}\;. \label{boundary phase from a reduction for general N}
\end{align}
The octahedral gluing structure (including the internal edges $\{C_* \}$) can be read off by carefully drawing the decomposition of $M$  using $\frac{k}6 N(N^2-1)$ octahedra.
The Lagrangian $\CL_N (\Delta \textrm{ or } M$) can be constructed as the image of a product of octahedron Lagrangian $\CL(\Diamond) = \{ e^{Z''}+e^{-Z}-1=0\}$ under the corresponding symplectic reduction.   In  section \ref{sec: PGL(N) gluing for 4_1}, we will illustrate the idea with the figure-eight knot complement as an example.
Since the basic quantum geometrical structure of an octahedron is equivalent to that of a $PGL(2)$ tetrahedron, the $PGL(N)$ CS ptn on $M$
can be computed using the recipe  \eqref{step 1 for PGL(2) CS}, \eqref{step 2 for PGL(2) CS}, \eqref{step 3 for PGL(2) CS} for the $PGL(2)$ CS ptn except that
we replace the $k$ tetrahedra by $\frac{k}6 N(N^2-1)$ octahedra accompanied by
the gluing rules.

Let's focus on the case when $M$ is a hyperbolic knot complement. When the knot complement can be decomposed into $k$ tetrahedra, $\frac{k}6 N(N^2-1)$ octahedra are necessary for the $N$-decomposition and  there are same number of internal edges in the octahedral decomposition. Among these internal edges, $N-1$ of them are not linearly independent. The remaining $N-1$ correspond to the boundary phase space of $M$ whose dimension is $2(N-1)$ parametrized by  $PGL(N)$ holomonies around the two cycles, longitude and meridian, at the boundary.
\begin{align}
\CP_N (\partial M) &= \left( \prod_{i=1}^k \prod_{(a,b,c,d)}\CP(\partial \Diamond^{(i)}_{(a,b,c,d)}) \right)//\{C_{*}=0 \}\;. \nn
\\
& =\{ {\bf m}:=\textrm{Hol}_{\textrm{meridian}} (\CA),\;{\bf l}:=\textrm{Hol}_{\textrm{longitude}} (\CA) \}/PGL(N)\;.
\label{symplectic reduction for knot complement}
\end{align}
Classically, the matrices $\bf{m}$ and $\bf{l}$ commute.
For quantization, we choose a polarization of the boundary phase space such that  meridian variables are positions  and longitudinal variables are momenta.
In this polarization, the $PGL(N)$ CS ptn  depends on $N-1$ meridian variables $\{\mu_i\}$ which parametrize $\bf{m}$ as in the eq.~\eqref{boundary condition}.
Using \eqref{step 1 for PGL(2) CS}\eqref{step 2 for PGL(2) CS}\eqref{step 3 for PGL(2) CS}, the $PGL(N)$ CS ptn  at  parabolic meridian ($\mu_i=0$) can be written as  ($\CM_N:= \frac{k}6 N (N^2-1)$)
\begin{align}
&Z^{\rm CS}_N (M)(\mu_i=0) \nn
\\
&=  \frac{e^{\frac{1}{2\hbar}(i \pi + \frac{\hbar}2)^2 f_N B_N^{-1} \nu_N}}{\sqrt{\det B_N}} \int \frac{\prod_{i=1}^k \prod_{(a,b,c,d)} d X^{(i)}_{a,b,c,d}}{(2\pi i \hbar)^{\CM_N/2}} \nn
\\
&\quad \times \exp \bigg{[}-\frac{1}{\hbar} ( i\pi
+\frac{\hbar}2) {\bf X} \cdot B_N^{-1} \nu_N + \frac{1}{2\hbar } {\bf X} \cdot B_N^{-1} A_N  \cdot {\bf X} \bigg{]}  \left( \prod_{i=1}^k \prod_{(a,b,c,d)} \psi_{\hbar}(X^{(i)}_{a,b,c,d}) \right)\;. \label{CS ptn at conformal point}
\end{align}
This is the main formula for the state-integral model which will be used rest part of this paper. By re-scaling the integral variables ${\bf X} $ to $b{\bf X}$, the symmetry $b\leftrightarrow b^{-1}$ becomes manifest using the $S$-duality property of the QDL \eqref{S-duality for QDL}. For later use, we write down the integral using
 `combinatorial flattenings' $(f,f',f'')_N$. 
 They are three integers associated to each octahedron satisfying
\begin{align}
A_N\cdot f_N+B_N \cdot f''_N=\nu_N\; \textrm{ and } (f_N)^{(i)}_{a,b,c,d}+(f'_N)^{(i)}_{a,b,c,d}+(f''_N)^{(i)}_{a,b,c,d}=1\;. \label{flattening}
\end{align}
These equations do not uniquely determine the flattenings but the ambiguity only affects a pre-factor in the state-integral of the form \eqref{ambiguity in state-integral} which is irrelevant in our discussion.
To write down the state-integral model we need to know the square-matrices $(A_N,B_N)$ of size $\CM_N$ and an $\CM_N$-column $\nu_N$. In section \ref{sec: PGL(N) gluing for 4_1}, these datum will be explicitly constructed for figure-eight knot complement $M$. For more knot (or link) complements, the $PGL(N)$ gluing datum is available in a recent version of {\tt SnapPy} \cite{SnapPy} up to $N=15$.

%%%%%%%%%%%%%%%%%%%%%%%%%%%%%
\subsection{Perturbative CS invariants from the state-integral model } 

\label{sec: perturbative expansion}

Using the method of steepest descent, the asymptotic expansion of the state-integral \eqref{CS ptn at conformal point} in the small $\hbar$ limit can be studied. We use the following asymptotic expansion of the QDL \eqref{Q.D.L}, \footnote{The notation $\widetilde{\rm Li}_{k}(e^{-X})$ is misleading. 
They are really functions of $X$ rather than of $e^{-X}$ \cite{Dimofte:2012qj}. }
\begin{align}
&\log \psi_{\hbar} (X)  =   \sum_{n=0}^\infty \frac{B_n \hbar^{n-1}}{n!}\widetilde{\rm Li}_{2-n}(e^{-X})\;, \textrm{ as } \hbar \rightarrow 0 \;, \label{asymptotic expansion of Q.D.L}
\end{align}
where $B_n$ is $n$-th Bernoulli's number ($B_0=1, B_1= \half,\ldots$).
The functions $\widetilde{\rm Li}_k$ with non-negative $k$ are defined by polylogarithm functions ${\rm Li}_k$
\begin{align}
\widetilde{\rm Li}_{k}(e^{-X}) ={\rm Li}_k (e^{-X})\;, \quad k=0,-1,-2,\ldots .
\end{align}
which are entire functions on $X$. The functions $\widetilde{\rm Li}_{k=1,2}(X)$ are equal to 
${\rm Li}_{k=1,2} (e^{-X})$ with the standard choice of branch-cuts when $0<\textrm{Im}(X)<2\pi$ or $\textrm{Re}(X)>0$. For $\textrm{Re}(X)$, 
$\widetilde{\rm Li}_{k=1,2}$ are obtained from ${\rm Li}_{k=1,2} (e^{-X})$ 
by aligning the branch-cuts along the imaginary axis. 
The branch-cuts are illustrated in Figure 9 of \cite{Dimofte:2012qj}. 
For practical purposes, one use the following relations to evaluate $\widetilde{\rm Li}_{k=1,2}$:
\begin{align}
&\widetilde{\rm Li}_{1}(e^{-X}) - {\rm Li}_1 (e^{-X}) = \left\{
                                                    \begin{array}{ll}
                                                      2\pi i \big{[}\frac{\textrm{Im}(X)}{(2\pi)}\big{]}-2\pi^2 \big{[}\frac{\textrm{Im}(X)}{(2\pi)}\big{]} \big{(}\big{[}\frac{\textrm{Im}(X)}{(2\pi)}\big{]}+1 \big{)} , \quad & \hbox{if $\textrm{Re}(X)<0$}\;,  \\
                                                      0, & \hbox{\hbox{if $\textrm{Re}(X)>0$}\;.}
                                                    \end{array}
                                                  \right. \nn
\\
&\widetilde{\rm Li}_{2}(e^{-X}) - {\rm Li}_2 (e^{-X}) = \left\{
                                                    \begin{array}{ll}
                                                      -2\pi i X \big{[}\frac{\textrm{Im}(X)}{(2\pi)}\big{]} , \quad & \hbox{if $\textrm{Re}(X)<0$}\;, \\
                                                      0, & \hbox{\hbox{if $\textrm{Re}(X)>0$}\;.}
                                                    \end{array}
                                                  \right.
\end{align}
Here $[x]$ denotes the floor of $x$, e.g. $[3/2]=1$. Physically, the branch-cut for $\widetilde{\rm Li}_{1,2}(e^{-X})$ comes from colliding poles and zeros of the QDL  \eqref{poles and zeros of QDL} when $\hbar = 2 \pi i b^2 \rightarrow 0$ with real $b$.
In the limit $\hbar \rightarrow 0$, the saddle point equations for the state-integral are
\begin{align}
A_N \cdot {\bf X} + B_N \cdot {\bf X}'' = i \pi \nu_N\;. \label{saddle point e.o.m}
\end{align}
where $X'':=-\widetilde{\rm Li}_1 (X)$.  Restricted on $0<\textrm{Im}(X)<\pi$, the  equations are equivalent to the gluing equations for the vertex variables of the octahedra
in the $N$-decomposition:
\begin{align}
&\textrm{internal vertex conditions : }C_* (\{Z,Z',Z''\}) =0\;, \nn
\\
&\textrm{meridian conditions : }\mu_i (\{Z,Z',Z''\})=0\;,  \nn
\\
&\qquad \textrm{with }Z+Z'+Z''=i\pi\quad \textrm{and} \quad e^{Z''}+e^{-Z}-1=0 \;.\label{gluing equations}
\end{align}
The  vertex variables  $\{Z,Z',Z''\}$ satisfying the gluing equations can be  mapped  to a saddle point ${\bf X}$ as follows ($X':=i \pi  - X-X''$)
\begin{align}
&(X,X',X'')_\g= (Z,Z',Z'')_\g \;\quad \textrm{if $(X_\Diamond,P_\Diamond)_\g = (Z,Z'')_\g$}\;, \nn
\\
&(X,X',X'')_\g= (Z',Z'',Z)_\g \;\quad \textrm{if $(X_\Diamond,P_\Diamond)_\g= (Z',Z)_\g$}\;, \nn
\\
&(X,X',X'')_\g= (Z'',Z,Z')_\g\;\quad \textrm{if $(X_\Diamond,P_\Diamond)_\g= (Z'',Z')_\g$}\;. \label{integral variables   and vertex variables}
\end{align}
The index $\gamma=1,\ldots, \CM_N$ labels  $\CM_N$ octahedra $\Diamond^{(i)}_{(a,b,c,d)}$ in the $N$-decomposition. 
The perturbative expansion of the state-integral  can be written as
\begin{align}
&Z^{CS}_N(M;\a) \simeq  \frac{1}{\hbar^{3/2}}\exp \left( \frac{1}{\hbar} S_0^{(\a)} + S_1^{(\a)}+\ldots \hbar^{n-1}S_n^{(\a)}+\ldots\right)\;, \quad \textrm{as $\hbar\rightarrow0$} \;.
\end{align}
Here $Z^{CS}_{N}(M;\a)$ is the state-integral \eqref{CS ptn at conformal point} along the Lefschetz thimble $\mathcal{J}_\a$ associated to a saddle point $\mathbf{X}^{(\a)}$. Schematically, the state-integral is of the form  
\begin{align}
\int d^{\CM_N
}\mathbf{X} \;e^{\CI(\mathbf{X})}\;.
\end{align}
We denote the real part of $\CI(\mathbf{X})$ by $h(\mathbf{X})$ and  consider the downward flow equations,
\begin{align}
\frac{d\mathbf{X}}{dt} = - \frac{\partial h}{\partial \overline{\mathbf{X}}}\;, \quad \frac{d\overline{\mathbf{X}}}{dt} = - \frac{\partial h}{\partial \mathbf{X}}\;.
\end{align}
Similarly, upward flow equations  can be defined by reversing the signs in the above. Along the downward  (upward) flow, the real part $h$ always  decreases (increases) while the imaginary part $\textrm{Im}(\CI)$ remains constant.  The $\CJ_{\a}$ is a set  of points in $\mathbb{C}^{\CM_N}$  that can be reached  at any $t$ by a downward flow starting from $\mathbf{X}^{(\a)}$ at $t=-\infty$. It defines a middle dimensional contour in $\mathbb{C}^{\CM_N}$ satisfying the two conditions:
\begin{align}
&1.\; \textrm{The phase  of the integrand stays constant along $\mathcal{J}_\a$}\;, \nn
\\
&2.\; X^{(\a)} \in \mathcal{J}_\a  
\mbox{ maximizes the absolute value of the integrand along } 
\mathcal{J}_\a \;.
\label{Good contour}
\end{align}
The integration along  Lefschetz thimbles $\CJ_{\a}$ are always convergent and they provide a basis of convergent contour. For any convergent contour $\mathcal{C}$,
\begin{align}
\mathcal{C}= \sum_\a m_\a \CJ_{\a} \;,\textrm{ which means } \int_{\mathcal{C}} d\mathbf{X} \;e^{\CI} = \sum_{\a}m_\a\int_{\CJ_\a} d\mathbf{X} \;e^{\CI}\;.
\end{align}
The coefficients $m_\a$ can be determined by counting upward flows that start from $\mathbf{X}^{(\a)}$ to $\mathcal{C}$ \cite{Witten:2010cx}. 
The perturbative expansion coefficients $\{ S_n^{(\a)}\}$ can be computed using a saddle point approximation \cite{Dimofte:2012qj}
\begin{align}
&S^{(\a)}_0 =-\frac{1}2 ({\bf X}-i \pi f_N)\cdot  ({\bf X}''+i \pi f''_N)  +\sum_{i=1}^\CM {\rm Li}_2 (e^{-X_i}) |_{X=X^{(\a)}} \;, \nn
\\
&S^{(\a)}_1 = - \frac{1}2 \log \left(  (\prod_\gamma x_\g^{f''_\g}(x_\g'')^{-f_\g})\det (A_N \cdot \Delta_{x''}+B_N \cdot \Delta_{x^{-1}}) \right)|_{X=X^{(\a)}}\;,\nn
\\
&S^{(\a)}_2 =\frac{1}8 \Gamma^{(4)}_\g(\Pi_{\g\g})^2+\frac{1}8 \Pi_{\g \g}\Gamma^{(3)}_\g \Pi_{\g \d}\Gamma_\d^{(3)} \Pi_{\d\d} +\frac{1}{12} \Gamma_\g^{(3)} (\Pi_{\g\d})^3 \Gamma_\d^{(3)}+ \half \Gamma_\g^{(1)} \Pi_{\g\d} \Gamma^{(3)}_\d \Pi_{\d\d}  \nn
\\
&\qquad + \half \Gamma^{(2)}\Pi_{\g\g}+ \frac{1}2 \Gamma^{(1)}_\g \Pi_{\g \d}\Gamma^{(1)}_{\d} + \Gamma^{(0)}|_{X=X^{(\a)}}\;, \nn
\\
&S_3^{(\a)} = (\mbox{see Figure 2 and 3 of \cite{Dimofte:2012qj}}).
\label{perturbative invariants}
\end{align}
Here $\Delta_{x''}:=\textrm{diag}\{x''_1, x''_2, \ldots, x''_{\CM_N} \}$ and $\Delta_{x^{-1}}:=\textrm{diag}\{x^{-1}_1, \ldots , x^{-1}_{\CM_N} \}$ with $x:=e^{X}, x':=e^{X'} = (1-x)^{-1}$ and  $x'':=e^{X''}=(1-x^{-1})$.
Summation over repeated  indices ($\beta, \gamma$) are assumed.
Propagator and interaction vertices from the state-integral \eqref{CS ptn at conformal point} are
\begin{align}
&\Pi := (-B_N^{-1}\cdot A_N+\Delta_{x'})^{-1}\;, \quad \textrm{(propagator)} \nn
\\
&\Gamma^{(0)}=\frac{1}8 f^T_N B_N^{-1}A_N f_N - \frac{1}{12}\sum_\g x'_\g\;, \quad \Gamma_\g^{(1)}:=\frac{x'_\g-(B_N^{-1}\nu_N)_\g}2\;, \quad \Gamma_\g^{(2)}:=\half x_\g (x'_\g)^2\;, \nn
\\
&\Gamma_\g^{(3)}:=- x_\g (x'_\g)^2 \;, \quad \Gamma_\g^{(4)}:=-x_\g (1+x_\g)(x'_\g)^3 \;,\quad (\textrm{vertices})\;.
\end{align}
Higher  invariants $\{S^{(\a)}_n\}_{n\ge 3}$ can also be expressed  in terms of  generalized Neunmann-Zagier datum $\{ A_N, B_N, {\bf X}^{(\a)},f_N, f'_N, f''_N\}$ using the Feynman rules in  \cite{Dimofte:2012qj}.  For example, we need to consider $40$ Feynmann diagrams depicted in Figure 1,2 and 3 in \cite{Dimofte:2012qj} to compute the 3-loop invariant $S_3$.
In general, there are several  saddle points ${\bf X}^{(\a)}$ satisfying \eqref{saddle point e.o.m} and from the saddle points ${\bf X}^{(\a)}$,  
$PGL(N)$ flat connections $\CA^{(\a)}$  can be constructed \cite{2011arXiv1111.2828G,2012arXiv1207.6711G,Dimofte:2013iv,Garoufalidis:2013upa}.  
Under the identification   \eqref{integral variables   and vertex variables}, a saddle point ${\bf X}^{({\rm conj})}$ corresponding to $\CA^{({\rm conj})}$
 \eqref{def of A(geom) and A(conj)} is characterized by two properties:
\begin{align}
&1. \textrm{ Vertex variables $(Z,Z',Z'')$ are constant on octahedra in each tetrahedron,}  \nn
\\
&\quad \textrm{i.e. $Z^{(i)}_{(a,b,c,d)}= Z^{(j)}_{(a',b',c',d')} $ if $i=j$}\;.  \nn
\\
&2. \textrm{ $0< \textrm{Im}(Z_\g),\textrm{Im}(Z'_\g),\textrm{Im}(Z''_\g)<\pi$}\;, \; \textrm{for all } \gamma. \label{Saddle point X(conj)}
\end{align}
The first property is true for every saddle points ${\bf X}^{(\a)}$ whose corresponding  $PGL(N)$ flat connections $\CA^{(\a)}$  can be constructed 
by embedding a $PGL(2)$ flat connection through the $N$-dimensional irreducible representation of $PGL(2)$. 
By imposing these two conditions, the saddle point equations are reduced to the gluing equations \eqref{gluing equations} for $N=2$. 
A solution to the these gluing equations for $N=2$ is called positive angle structure and known to give a complete hyperbolic structure on a knot complement $M$.\footnote{The gluing equations describe how to glue ideal tetrahedra without any conical  singularity by tuning the shape (edge variable $Z,Z',Z''$) of each tetrahedron to form the 3-manifold $M$. The conditions $0<\textrm{Im}(Z,Z',Z'')<\pi, Z+Z'+Z''=i \pi$ and $e^{Z''}+e^{-Z}-1=0$ are necessary for an ideal tetrahedron to be embedded in the hyperbolic space $H^3$.  Since each ideal tetrahedron in $H^3$ has a hyperbolic structure, a solution to the gluing equations defines a smooth hyperbolic structure on $M$. Additional meridian condition guarantees the completeness of the hyperbolic metric. }  Mostow's rigidity theorem  guarantee the uniqueness of the positive angle structure for hyperbolic knot complements $M$ if it exists. Existence of the structure depends on the ideal triangulation  of the 3-manifold and we will always use a triangulation $\CT$ which admits the structure.   
Among saddle points, the saddle point ${\bf X}^{(\textrm{conj})}$ minimizes the imaginary part of $S_0$.
\begin{align}
\textrm{Im}[S_0^{(\textrm{conj})}]\leq \textrm{Im}[S_0^{(\a)} ]\;, \quad \textrm{for any $(\a)$}\;.
\end{align}
To prove the assumption  \eqref{Two things on CS ptn-1} using the state-integral model, it is necessary and sufficient to show that 
\begin{align}
m_{\alpha} \neq 0 \textrm{ if and only if } (\alpha)=(\textrm{conj})\;, \textrm{ where $\mathcal{C}_\mathbb{R}  =\sum_\a m_{\a} \CJ_{\a}$}\;. \label{decomposition of C_R}
\end{align}
For  some simplest cases, the  upward flow from $\mathbf{X}^{(\textrm{conj})}$ to $\mathcal{C}_{\mathbb{R}}$ can be explicitly constructed and it can be shown that $m_\a =0$ for $(\a) \neq  (\textrm{conj})$. See  appendix  \ref{sec : flow analysis}. If a upward flow connecting $\CA^{(\textrm{conj})}$ to $\mathcal{C}_{\mathbb{R}}$ for $N=2$ is constructed, then  upward flow for general $N$ can be  constructed as $(Z,Z',Z'')^{(i)}_{(a,b,c,d)}(t) =(Z,Z',Z'')^{(i)}_{N=2}(t)$. 
The upward flow analysis gets much harder as the number of integral variables increases and we leave the  proof of \eqref{decomposition of C_R} for general cases as future problem.  

\subsection{Numerical checks for the conjecture \eqref{conjecture}  }

\subsubsection{Figure-eight knot complement}
\label{sec: PGL(N) gluing for 4_1}
\begin{figure}[h!]
\begin{center}
   \includegraphics[width=1\textwidth]{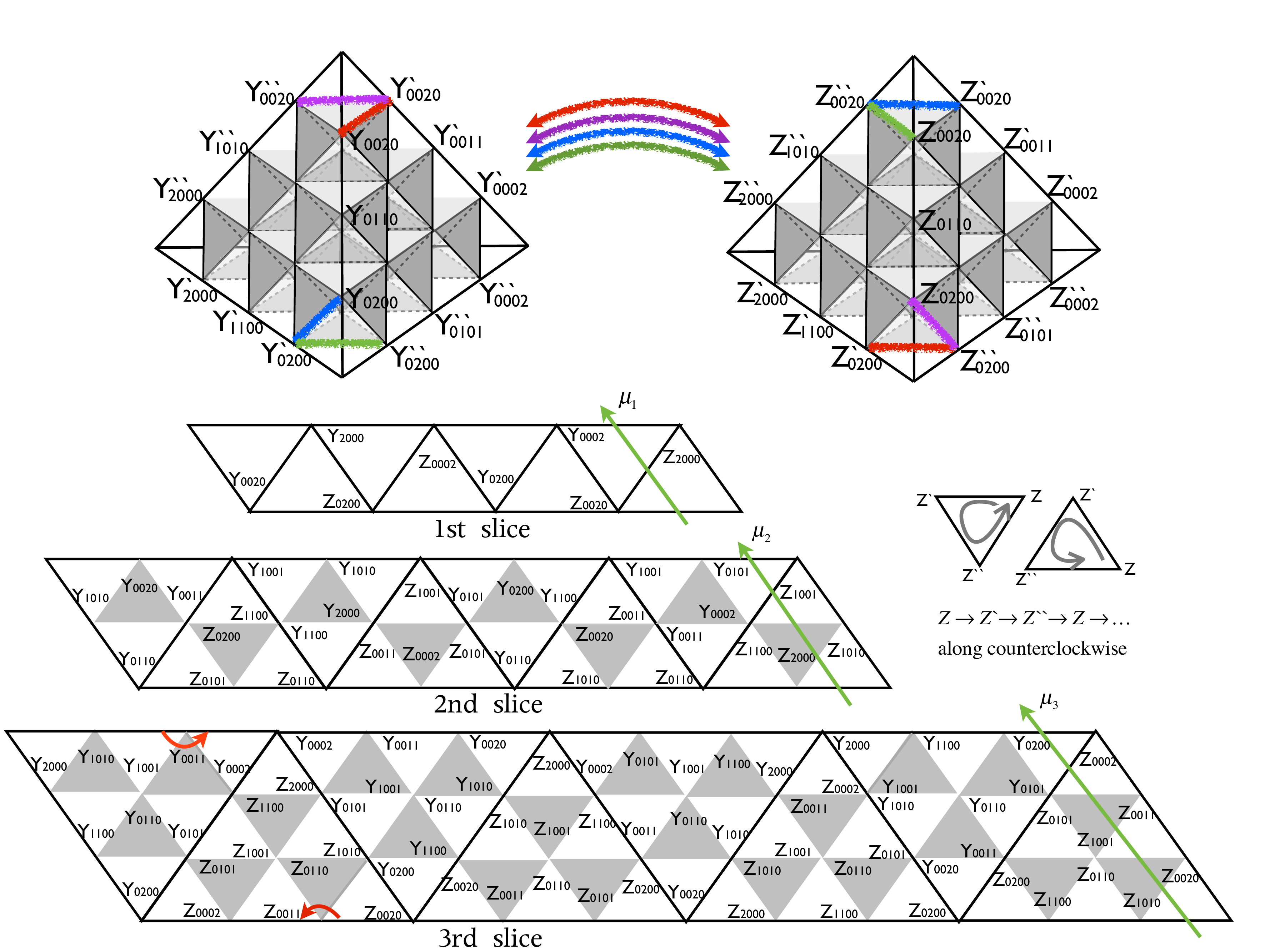}
   \end{center}
   \caption{$N$-decomposition of figure-eight knot complement when $N=4$. Three slices of  the torus boundary are drawn. In each slice,  horizontal and vertical directions are periodic. From the  diagram,  internal vertices and meridian variables $\{\mu_i\}_{i=1}^{3}$ can be read. For example, the internal vertex in red color at the 3rd slice is $C_{\circ}=Z''_{0,0,2,0}+Z'_{0,1,1,0}+Z_{0,0,1,1}+Y'_{1,0,0,1}+Y_{0,0,1,1}+Y''_{0,0,0,2}-2\pi i$. }
    \label{fig:N-decomposition for 4_1}
\end{figure}

The figure-eight knot complement, $M=S^3\backslash \mathbf{4}_1$,  can be decomposed into two ideal tetrahedra as depicted in Figure \ref{fig:N-decomposition for 4_1}. Decomposing each tetrahedron into a pyramid of $\frac{1}6 N (N^2-1)$ octahedra, we obtain the $N$-decomposition of $M$. The vertex variables of the octahedra in the first tetrahedron are denoted by $Z^{(1)}_{(a,b,c,d)}=Y_{(a,b,c,d)}$ and the other one  by $Z^{(2)}_{(a,b,c,d)}:=Z_{(a,b,c,d)}$. For $N=4$, the octahedral decomposition is depicted in Figure \ref{fig:N-decomposition for 4_1}.
In general there are three types of internal vertices:
\begin{itemize}

\item ``Edge'' type : located on edges of tetrahedra ,

\item ``Face'' type : located on faces of  tetrahedra ,

\item ``Interior'' type : located inside tetrahedra .

\end{itemize}
Internal vertices of edge and face type depend on the gluing data \eqref{tetrahedral decom of M} of tetrahedra while internal vertices of interior type do not.

\vskip 0.5cm 

For $N=2$, there are two internal vertices
\begin{align}
 &C_1 =Y_{0, 0, 0, 0} + 2 Y'_{0, 0, 0, 0}+
 Z_{0, 0, 0, 0} + 2 Z'_{0, 0, 0, 0}- 2\pi i \;,\nn
 \\
 &C_2=Y_{0, 0, 0, 0} + 2 Y''_{0, 0, 0, 0} + Z_{0, 0, 0, 0} +
 2 Z''_{0, 0, 0, 0}- 2 \pi i \;. \nn
\end{align}
Both of them are of edge type and there is no internal vertices of face or interior type. The two vertices are linearly dependent, $C_1 +C_2 = 0$. The single meridian variable is
\begin{align}
\mu  = Z'_{0,0,0,0}- Y''_{0,0,0,0}\;. \nn
\end{align}
We choose  the position variables $\widetilde{\bf X}$ in \eqref{Sp(2k,Z)+affine shifts} to be $\widetilde{\bf X}=\{C_2 , \mu \}$ and the polarization of  each octahedron to be $\Pi_{{\bf X}_\Diamond}=({\bf X}_{\Diamond} =\{ Y'_{0,0,0,},Z_{0,0,0,0}\},{\bf P}_{\Diamond} =\{ Y_{0,0,0,},Z''_{0,0,0,0}\})$. Then
the data $A, B$ and $\nu$ in \eqref{Sp(2k,Z)+affine shifts} are
\begin{align}
A_{N=2} = \left(\begin{array}{cc}-2 & 1 \\ 1 & -1\end{array}\right)\;, \quad B_{N=2}= \left(\begin{array}{cc}-1 & 2 \\ 1 & -1\end{array}\right) \;, \quad \nu_{N=2} = \left(\begin{array}{c}0 \\0\end{array}\right)\;.
\end{align}

\vskip 0.5cm

For $N=3$, the octahedral gluing equations are studied in sec 7.4. of \cite{Dimofte:2011ju}. There are 8 internal vertices (no interior type)
\begin{align}
&\textrm{Edge type : }&Y_{0, 0, 0, 1} + Y'_{0, 0, 0, 1} +
  Y'_{0, 1, 0, 0} + Z_{0, 0, 1, 0} + Z'_{0, 0, 1, 0} +
  Z'_{1, 0, 0, 0}-2\pi i\;, \nn
  \\
  && Y_{1, 0, 0, 0} +
  Y'_{0, 0, 1, 0} + Y'_{1, 0, 0, 0} + Z_{0, 1, 0, 0} + Z'_{0, 0, 0, 1} +
  Z'_{0, 1, 0, 0}-2\pi i\;, \nn
  \\
  &&Y_{0, 1, 0, 0} +
  Y''_{0, 1, 0, 0} + Y''_{1, 0, 0, 0} + Z_{0, 0, 0, 1} +
  Z''_{0, 0, 0, 1} + Z''_{0, 0, 1, 0}-2\pi i\;, \nn
  \\
  &&Y_{0, 0, 1, 0} + Y''_{0, 0, 0, 1} + Y''_{0, 0, 1, 0} + Z_{1, 0, 0, 0} + Z''_{0, 1, 0, 0} + Z''_{1, 0, 0, 0}-2\pi i\;.
\end{align}
\begin{align}
&\textrm{Face type : }& Y_{0, 0, 1, 0} + Y'_{1, 0, 0, 0} + Y''_{0, 0, 0, 1} + Z_{0, 0, 0, 1} +
 Z'_{0, 1, 0, 0} + Z''_{0, 0, 1, 0}-2\pi i\;, \nn
  \\
  &&Y_{0, 1, 0, 0} + Y'_{0, 0, 0, 1} + Y''_{1, 0, 0, 0} + Z_{1, 0, 0, 0} +
 Z'_{0, 0, 1, 0} + Z''_{0, 1, 0, 0}-2\pi i\;, \nn
  \\
  &&Y_{1, 0, 0, 0} + Y'_{0, 0, 1, 0} + Y''_{0, 1, 0, 0} + Z_{0, 0, 1, 0} +
 Z'_{1, 0, 0, 0} + Z''_{0, 0, 0, 1}-2 \pi i\;, \nn
  \\
  &&Y_{0, 0, 0, 1} + Y'_{0, 1, 0, 0} + Y''_{0, 0, 1, 0} + Z_{0, 1, 0, 0} +
 Z'_{0, 0, 0, 1} + Z''_{1, 0, 0, 0}-2\pi i\;.
\end{align}
The two meridian variables are
\begin{align}
&\mu_1 = -Y''_{0, 0, 0, 1} + Z'_{1, 0, 0, 0}\;, \nn
\\
&\mu_2 = -Y''_{0, 1, 0, 0} + Z'_{0, 0, 0, 1} + Z'_{0, 0, 1, 0} - Z'_{1, 0, 0, 0}\;.
\end{align}
We choose the position variables $\widetilde{\bf X}$ in \eqref{Sp(2k,Z)+affine shifts} as follows
\begin{align}
\widetilde{\bf X} = \big{\{} & \textrm{all internal vertices of edge type except the 1st one} \nn
\\
&\textrm{all  internal vertices of face type except the 1st one}\;,\nn
\\
&\textrm{meridian variables }\mu_1\; \textrm{and} \; \mu_2 \big{\}}\;.
\end{align}
One can see that all elements of $\widetilde{\bf X}$ are linearly independent.
In this choice, $(A_3, B_3, \nu_3)$ in \eqref{Sp(2k,Z)+affine shifts} are $\nu_3= \vec{0}$ and 
\begin{align}
A_3 =   \left(\begin{array}{cccccccc}
0& 1& 0& 1& 0& 0& -1& 0
\\
-1& 0& 0& -1& 0& 0& 1& 0
\\
0& -1& -1& 0& 0& 0& 0& 1
\\
0& 0& 1& -1& 0& -1& 0& 1
\\
-1& 1& 0& 0& 0& 1& 0& -1
\\
1& -1& 0& 0& 1& 0& -1& 0
\\
0& 0& 1& 0& 0& 0& 0& -1
\\
1& 0& 0& 0& 0& -1& -1& 1\end{array}\right)\;, \quad
B_3 =   \left(\begin{array}{cccccccc}
0& 0& 0& 1& -1& 0& -1& 0
\\
0& 0& 0& -1& 0& 1& 1& 0
\\0& 0& -1& 0& 1& 0& 0& 1
\\1& 0& 0& -1& 1& -1& 0& 0
\\-1& 0& 0& 1& 0& 0& 1& -1
\\0& -1& 1& 0& 0& 0& -1& 1
\\0& 0& 1& 0& 0& 0& 0& -1
\\1& 0& 0& 0& 0& -1& -1& 1\end{array}\right)\;.
\end{align}
Here, we choose 
${\bf X}_{\Diamond}= \{Y'_{0, 1, 0, 0}, Y'_{0, 0, 1, 0}, Y'_{0, 0, 0, 1}, Y'_{1, 0, 0, 0},
 Z_{0, 1, 0, 0}, Z_{0, 0, 1, 0}, Z_{0, 0, 0, 1}, Z_{1, 0, 0, 0}\}$ and   ${\bf P}_{\Diamond}= \{Y_{0, 1, 0, 0}, Y_{0, 0, 1, 0}, Y_{0, 0, 0, 1}, Y_{1, 0, 0, 0},
 Z''_{0, 1, 0, 0}, Z''_{0, 0, 1, 0}, Z''_{0, 0, 0, 1}, Z''_{1, 0, 0, 0}\}$.

\vskip 0.5cm

For $N=4$, there are 20 internal vertices.
\begin{align}
&\textrm{Edge type : }& Y_{0, 0, 0, 2} + Y'_{0, 0, 0, 2} +
  Y'_{0, 2, 0, 0} + Z_{0, 0, 2, 0} + Z'_{0, 0, 2, 0} +
  Z'_{2, 0, 0, 0}-2\pi i \;, \nn
  \\
  && Y_{1, 0, 0, 1} +
  Y'_{0, 0, 1, 1} + Y'_{1, 1, 0, 0} + Z_{0, 1, 1, 0} + Z'_{0, 0, 1, 1} +
  Z'_{1, 1, 0, 0}-2\pi i\;, \nn
  \\
  && Y_{2, 0, 0, 0} +
  Y'_{0, 0, 2, 0} + Y'_{2, 0, 0, 0} + Z_{0, 2, 0, 0} + Z'_{0, 0, 0, 2} +
  Z'_{0, 2, 0, 0}-2\pi i\;, \nn
  \\
  && Y_{0, 2, 0, 0} +
  Y''_{0, 2, 0, 0} + Y''_{2, 0, 0, 0} + Z_{0, 0, 0, 2} +
  Z''_{0, 0, 0, 2} + Z''_{0, 0, 2, 0}-2\pi i\;, \nn
  \\
  &&Y_{0, 1, 1, 0} + Y''_{0, 1, 0, 1} + Y''_{1, 0, 1, 0} + Z_{1, 0, 0, 1} +
  Z''_{0, 1, 0, 1} + Z''_{1, 0, 1, 0}-2\pi i\;, \nn
  \\
  &&Y_{0, 0, 2, 0} + Y''_{0, 0, 0, 2} + Y''_{0, 0, 2, 0} + Z_{2, 0, 0, 0} +
  Z''_{0, 2, 0, 0} + Z''_{2, 0, 0, 0}-2\pi i\;.
\end{align}
\begin{align}
&\textrm{Interior type :}& \; Z_{0, 1, 1, 0} + Z_{1, 0, 0, 1} + Z'_{0, 0, 1, 1} + Z'_{1, 1, 0, 0} +  Z''_{0, 1, 0, 1} + Z''_{1, 0, 1, 0}-2\pi i\;, \nn
  \\
  &&Y_{0, 1, 1, 0} + Y_{1, 0, 0, 1} + Y'_{0, 0, 1, 1} + Y'_{1, 1, 0, 0} +  Y''_{0, 1, 0, 1} + Y''_{1, 0, 1, 0}-2\pi i\;.
\end{align}
\begin{align}
&\textrm{Face type : }& Y_{0, 0, 1, 1} + Y'_{1, 0, 0, 1} + Y''_{0, 0, 0, 2} + Z_{0, 0, 1, 1} +  Z'_{0, 1, 1, 0} + Z''_{0, 0, 2, 0}-2\pi i\;, \nn
  \\
  && Y_{1, 0, 1, 0} + Y'_{2, 0, 0, 0} + Y''_{1, 0, 0, 1} + Z_{0, 0, 0, 2} +   Z'_{0, 1, 0, 1} + Z''_{0, 0, 1, 1}-2\pi i\;, \nn
  \\
  && Y_{0, 0, 2, 0} + Y'_{1, 0, 1, 0} + Y''_{0, 0, 1, 1} + Z_{0, 1, 0, 1} +   Z'_{0, 2, 0, 0} + Z''_{0, 1, 1, 0}-2\pi i\;, \nn
  \\
  && Y_{0, 2, 0, 0} + Y'_{0, 1, 0, 1} + Y''_{1, 1, 0, 0} + Z_{1, 0, 1, 0} +   Z'_{0, 0, 2, 0} + Z''_{0, 1, 1, 0}-2\pi i\;, \nn
  \\
  && Y_{1, 1, 0, 0} + Y'_{1, 0, 0, 1} + Y''_{2, 0, 0, 0} + Z_{1, 1, 0, 0} +   Z'_{0, 1, 1, 0} + Z''_{0, 2, 0, 0}-2\pi i\;, \nn
  \\
  && Y_{0, 1, 0, 1} + Y'_{0, 0, 0, 2} + Y''_{1, 0, 0, 1} + Z_{2, 0, 0, 0} +   Z'_{1, 0, 1, 0} + Z''_{1, 1, 0, 0}-2\pi i\;, \nn
  \\
  && Y_{1, 0, 1, 0} + Y'_{0, 0, 2, 0} + Y''_{0, 1, 1, 0} + Z_{1, 0, 1, 0} +   Z'_{2, 0, 0, 0} + Z''_{1, 0, 0, 1}-2\pi i\;, \nn
  \\
  && Y_{1, 1, 0, 0} + Y'_{0, 1, 1, 0} + Y''_{0, 2, 0, 0} + Z_{0, 0, 2, 0} +   Z'_{1, 0, 1, 0} + Z''_{0, 0, 1, 1}-2\pi i\;, \nn
  \\
  &&  Y_{2, 0, 0, 0} + Y'_{1, 0, 1, 0} + Y''_{1, 1, 0, 0} + Z_{0, 0, 1, 1} +   Z'_{1, 0, 0, 1} + Z''_{0, 0, 0, 2}-2\pi i \;, \nn
  \\
  &&  Y_{0, 0, 1, 1} + Y'_{0, 1, 1, 0} + Y''_{0, 0, 2, 0} + Z_{0, 2, 0, 0} +   Z'_{0, 1, 0, 1} + Z''_{1, 1, 0, 0}-2\pi i\;, \nn
  \\
  && Y_{0, 0, 0, 2} + Y'_{0, 1, 0, 1} + Y''_{0, 0, 1, 1} + Z_{1, 1, 0, 0} +   Z'_{1, 0, 0, 1} + Z''_{2, 0, 0, 0}-2\pi i\;, \nn
  \\
  && Y_{0, 1, 0, 1} + Y'_{0, 2, 0, 0} + Y''_{0, 1, 1, 0} + Z_{0, 1, 0, 1} +   Z'_{0, 0, 0, 2} + Z''_{1, 0, 0, 1}-2\pi i\;.
\end{align}
The three meridian variables are (see Figure \ref{fig:N-decomposition for 4_1})
\begin{align}
&\mu_1 = -Y''_{0, 0, 0, 2} + Z'_{2, 0, 0, 0}\;, \quad \mu_2= -Y''_{0, 1, 0, 1} + Z'_{1, 0, 0, 1} +  Z'_{1, 0, 1, 0} - Z'_{2, 0, 0, 0} \nn
\\
&\mu_3 =-Y''_{0, 2, 0, 0} + Z'_{0, 0, 0, 2} +  Z'_{0, 0, 1, 1} + Z'_{0, 0, 2, 0} - Z'_{1, 0, 0, 1} - Z'_{1, 0, 1, 0}\;.
\end{align}
We choose the position variables $\widetilde{\bf X}$ in \eqref{Sp(2k,Z)+affine shifts} as follows
\begin{align}
\widetilde{\bf X} = \big{\{} &\textrm{all internal vertices of edge type except the 1st and 2nd}\;, \nn
\\
&\textrm{all internal vertices of face type except the 1st\;,} \nn
\\
&\textrm{all  internal vertices of interior type }\;, \nn
\\
&\textrm{meridian variables }\mu_1,\mu_2 \textrm{ and } \mu_3 \big{\}} \;.
\end{align}
One can check that the elements of $\widetilde{\bf X}$ are  linearly independent. The $20\times 20$ matrices $(A_4,B_4)$ and the vector $\nu_4$ can be straightforwardly obtained  with a  proper choice of $\Pi_{{\bf X}_\Diamond} =({\bf X}_{\Diamond},{\bf P}_{\Diamond})$.

\vskip 0.5cm

For general $N$, there are $\frac{1}3 N (N^2-1)$ internal vertices in the $N$-decomposition of figure-eight knot complement.

\vskip 0.5cm

\noindent
Edge type : $a=0,1,\ldots, {\rm max}:=N-2$
\begin{align}
&Z_{0, a, {\rm max} - a, 0} + Z'_{0, 0, {\rm max} - a, a} + Z'_{{\rm max} - a, a, 0, 0} +  Y'_{a, {\rm max} - a, 0, 0} + Y'_{0, 0, a, {\rm max} - a} + Y_{a, 0, 0, {\rm max} - a}-2\pi i, \nn
\\
&  Z''_{a, 0, {\rm max} - a, 0} + Z''_{0, a, 0, {\rm max} - a} + Z_{a, 0, 0, {\rm max} - a} +  Y''_{{\rm max} - a, 0, a, 0} + Y''_{0, {\rm max} - a, 0, a} + Y_{0, {\rm max} - a, a, 0} -2\pi i\;. \nn
\end{align}
Face type : $a,b\geq 0 ,\; a+b \leq {\rm max}-1$
\begin{align}
& Z_{0, a, {\rm max} - 1 - a - b, 1 + b} +    Z'_{0, 1 + a, {\rm max} - 1 - a - b, b} + Z''_{0, a, {\rm max} - a - b, b} \nn
\\
&\quad +Y_{b, 0, 1 + a, {\rm max} - 1 - a - b} +    Y'_{1 + b, 0, a, {\rm max} - 1 - a - b} + Y''_{b, 0, a, {\rm max} - a - b} -2\pi i\;, \nn
\\
& Z_{1 + a, b, {\rm max} - 1 - a - b, 0} +  Z''_{a, 1 + b, {\rm max} - 1 - a - b, 0} + Z'_{a, b, {\rm max} - a - b, 0}  \nn
\\
&\quad + Y_{b, {\rm max} - a - b, 0, a} + Y'_{b, {\rm max} - 1 - a - b, 0, 1 + a} +  Y''_{1 + b, {\rm max} - 1 - a - b, 0, a} -2\pi i \;,  \nn
\\
&Z_{{\rm max} - 1 - a - b, 0, 1 + b, a} + Z'_{{\rm max} - a - b, 0, b, a} +    Z''_{{\rm max} - 1 - a - b, 0, b, 1 + a} \nn
\\
&\quad +  Y_{1 + a, b, {\rm max} - 1 - a - b, 0} +  Y''_{a, 1 + b, {\rm max} - 1 - a - b, 0} + Y'_{a, b, {\rm max} - a - b, 0} -2\pi i\;, \nn
\\
&Z_{b, {\rm max} - a - b, 0, a} + Z'_{b, {\rm max} - 1 - a - b, 0, 1 + a} + Z''_{1 + b, {\rm max} - 1 - a - b, 0, a} \nn
\\
&\quad +  Y_{0, a, {\rm max} - 1 - a - b, 1 + b} +   Y'_{0, 1 + a, {\rm max} - 1 - a - b, b} + Y''_{0, a, {\rm max} - a - b, b}-2\pi i\;. \nn
\end{align}
Interior type : $a-1,b,c,d-1\geq 0,\; a+b+c+d={\rm max}$
\begin{align}
& Z_{a, b, c, d} + Z_{a - 1, b + 1, c + 1, d - 1} +   Z'_{a - 1, b, c + 1, d} + Z'_{a, b + 1, c, d - 1} +   Z''_{a, b, c + 1, d - 1} + Z''_{a - 1, b + 1, c, d} -2\pi i \;, \nn
\\
&Y_{a, b, c, d} + Y_{a - 1, b + 1, c + 1, d - 1} +   Y'_{a - 1, b, c + 1, d} + Y'_{a, b + 1, c, d - 1} +  Y''_{a, b, c + 1, d - 1} + Y''_{a - 1, b + 1, c, d}-2\pi i \;.\nn
\end{align}
There are $N-1$ meridian variables $\{\mu_i\}_{i=1}^{N-1}$
\begin{align}
\mu_i= \sum_{k=0}^{i-1}Z'_{N-1-i,0,i-1-k,k}-\sum_{k=0}^{i-2}Z'_{N-i,0,i-k-2,k}-
  Y''_{0, i-1, 0, N-1- i} \;.
\end{align}
We choose the position variables $\widetilde{X}$ in \eqref{Sp(2k,Z)+affine shifts} as follows
\begin{align}
\widetilde{\bf X} = \{ &\textrm{all internal vertices of edge type except   $\left[N/2 \right]$ entries },\nn
\\
&\textrm{all internal vertices of face type except   $\left[(N-1)/2\right]$ entries},\nn
\\
&\textrm{all internal vertices of interior type },\nn
\\
&\textrm{all meridian variables} \}\;. \nn
\end{align}
Here $[x]$ denote the floor of $x$, e.g. $[\frac{3}2]=1$. We need to carefully decide which  internal vertices should be  abandoned  in order to make the set $\widetilde{\bf X}$ linearly independent.  With a  choice of octahedron's polarization  $\Pi_{{\bf X}_\Diamond}=({\bf X}_\Diamond,{\bf P}_\Diamond)$, the datum $(A_N, B_N, \nu_N)$ in \eqref{Sp(2k,Z)+affine shifts}  can be straightforwardly calculated. One subtle thing is that $B_N$ is not invertible for a general choice of $\Pi_{\bf{X}_\Diamond}$. The state-integral  in \eqref{CS ptn at conformal point} make sense only when the  matrix $B_N$ is invertible. We need to carefully choose the octahedron's polarization $\Pi_{\bf{X}_\Diamond}$ such that $B_N$ is non-degenerate, which is always possible as shown  in  \cite{Dimofte:2012qj}.

The saddle point ${\bf X}^{(\rm conj)}$ satisfying the two conditions in \eqref{Saddle point X(conj)} and the flattenings $(f,f',f'')_N$  \eqref{flattening} are given by  (under the identification  \eqref{integral variables   and vertex variables})
\begin{align}
&Y_{a,b,c,d}=Y'_{a,b,c,d}=Y''_{a,b,c,d}= Z_{a,b,c,d}=Z'_{a,b,c,d}=Z''_{a,b,c,d} = i \pi/3\;, \nn
\\
&(f,f',f'')^{(Y)}_{(a,b,c,d)} =\left\{\begin{array}{c}(0,0,1)\quad \textrm{if $(X_\Diamond,P_\Diamond)^{(Y)}_{(a,b,c,d)} = (Y,Y'')_{(a,b,c,d)}$} \\ (0,1,0)\quad \textrm{if $(X_\Diamond,P_\Diamond)^{(Y)}_{(a,b,c,d)} = (Y',Y)_{(a,b,c,d)}$} \\
(1,0,0)\quad \textrm{if $(X_\Diamond,P_\Diamond)^{(Y)}_{(a,b,c,d)} = (Y'',Y')_{(a,b,c,d)}$}\end{array}\right.\;, \nn
\\
&(f,f',f'')^{(Z)}_{(a,b,c,d)} =\left\{\begin{array}{c}(0,1,0)\quad \textrm{if $(X_\Diamond,P_\Diamond)^{(Z)}_{(a,b,c,d)} = (Z,Z'')_{(a,b,c,d)}$} \\ (1,0,0)\quad \textrm{if $(X_\Diamond,P_\Diamond)^{(Z)}_{(a,b,c,d)} = (Z',Z)_{(a,b,c,d)}$} \\
(0,0,1)\quad \textrm{if $(X_\Diamond,P_\Diamond)^{(Z)}_{(a,b,c,d)} = (Z'',Z')_{(a,b,c,d)}$}\end{array}\right. \;.
\end{align}
From the  Neunmann-Zagier datum $\{ A_N, B_N, {\bf X}^{({\rm conj})},f_N, f'_N, f''_N\}$ for the $N$-decomposition, it is straightforward to compute  the perturbative invariants $\{S^{(\textrm{conj})}_n (N)\}$ for the figure-eight knot complement  using the formula in eq.~\eqref{perturbative invariants}. The classical part yields 
\begin{align}
&{\rm Im}[S^{(\textrm{conj})}_0(N)] ={\rm Im}[ \sum_{i=1}^{\CM_N} {\rm Li}_2 (e^{-X_i}) ] \nn
\\
&=  \textrm{Im} \big{[} \frac{1}{3} N (N^2-1){\rm Li}_2 (e^{- i \pi/3}) \big{]} = - \frac{1}{6} N (N^2-1) \textrm{vol} (S^3\backslash \mathbf{4}_1)  \;, \label{S0 for 4_1}
\end{align}
where we used the fact that $\textrm{vol} (S^3 \backslash \mathbf{4}_1)  =- 2 \textrm{ Im}({\rm Li}_2 (e^{-i \pi/3}))$. This  is compatible with  \eqref{classical action for geom,conj}. 
The one-loop invariants are
\begin{align}
&\textrm{Re}[S^{(\textrm{conj})}_1(N)]:=\textrm{Re}[- \frac{1}2 \log  \det \left(e^{i \pi /3}A_N + e^{-i \pi /3} B_N    \right) ]  \quad
\textrm{for $N=2,\ldots,30$} \nn
\\
&=\{-0.274653, -1.52226, -4.68107, -10.4071, -19.338, -32.13, -49.4353, -71.902,  \nn
\\
&\quad \;-100.178, -134.909, -176.745, -226.33, -284.312, -351.337, -428.0517, -515.10336,  \nn
\\
&\quad \;-613.1371, -722.7996,-844.7372,-979.5963,-1128.023 , -1290.6635,  -1468.1641, \nn
\\
&\quad  \; -1661.171, -1870.3305,-2096.2886, -2339.6916, -2601.1856, -2881.4169\} \nn
\end{align}
Their third-difference sequence ${\rm Re}[S^{'''}_1(N)]$ is \footnote{ $S_1^{'''}(N):=S_1^{''}(N+1)-S_1^{''}(N)$,\; $S_1^{''}(N):=S^{'}(N+1)-S^{'}(N)$ and $S^{'}_1(N):=S_1^{(\textrm{conj})}(N+1)-S_1^{(\textrm{conj})}(N)$.}
\begin{align}
&{\rm Re}[S^{'''}_1(N)] \quad  \textrm{for $N=2,\ldots,27$}  \nn
\\
& =\{-0.655958, -0.637856, -0.655893, -0.652562, -0.647830, -0.647560,  -0.647428, \nn \\
&\quad \;\; -0.647022,-0.646783, -0.646649, -0.646543,-0.646462,-0.646402, -0.646356, \nn
\\
&\quad \;\;  -0.646319, -0.646291,-0.646267,-0.646248,-0.646233, -0.646220, -0.646209,  \nn
\\
&\quad \;\; -0.646200, -0.646192, -0.646186, -0.646180, -0.646174\}\;. \nn
\end{align}
\begin{figure}[h!]
\begin{center}
   \includegraphics[width=.5\textwidth]{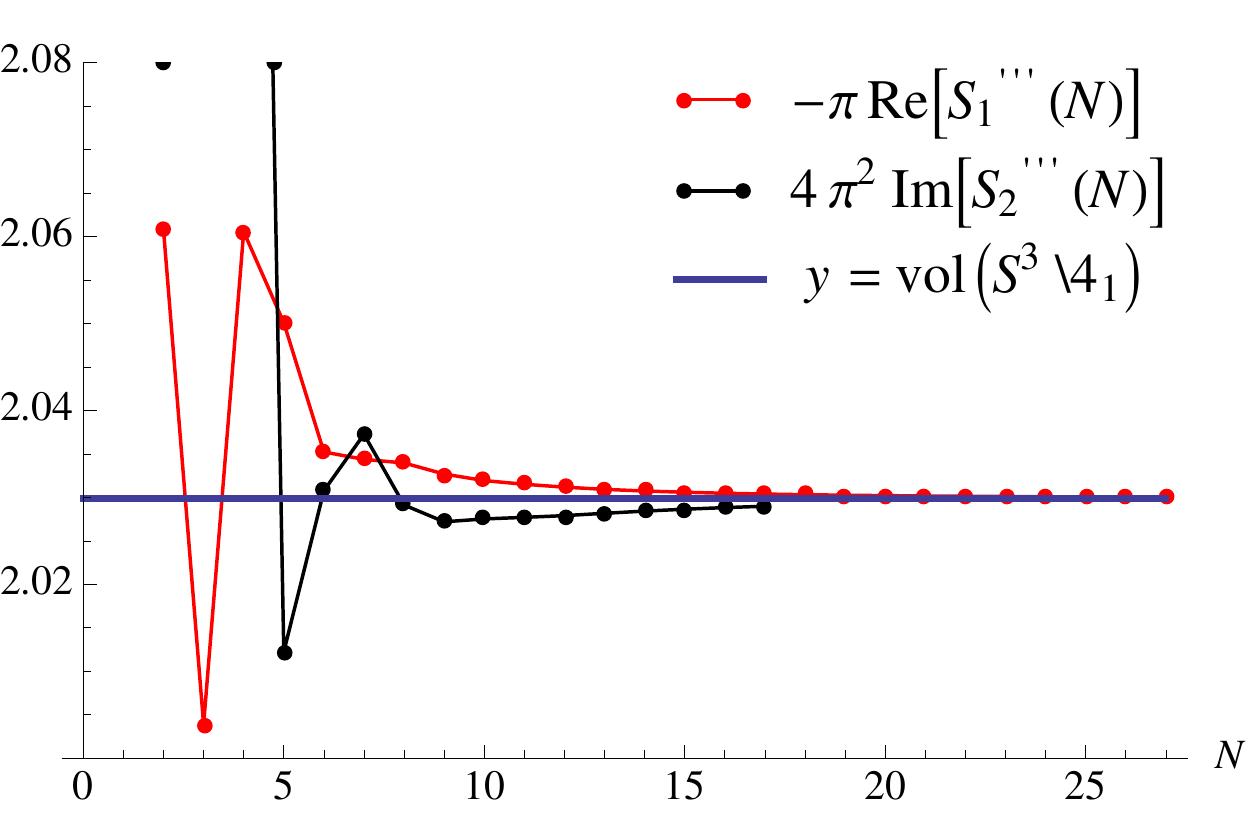}
   \end{center}
   \caption{As $N$ grows, both $-\pi {\rm Re}[S_1^{'''}(N)]$ and $4\pi^2{\rm Im}[S_2^{'''}(N)]$ quickly approach $\textrm{vol}(S^3\backslash\mathbf{4}_1) \approx 2.02988$. This gives a numerical evidence for the conjecture \eqref{conjecture} for $M=S^3\backslash \mathbf{4}_1$ when $n=1,2$.}
    \label{fig:Re[S1] and Im[S2]}
\end{figure}
Note that the  sequence rapidly converges to a  constant value $-0.6461...$ which is very close to $ -\frac{1}{\pi}\textrm{vol}(S^3\backslash\mathbf{4}_1) =-0.646132...$, as depicted in Figure \ref{fig:Re[S1] and Im[S2]}.
From this analysis, we numerically confirm that
\begin{align}
\textrm{Re}[S_1(N)] =  -\frac{N^3}{6 \pi } \textrm{vol}(S^3 \backslash\mathbf{4}_1) + (\textrm{sub-leading in $1/N$}) \quad \textrm{as } N\rightarrow \infty \;. \label{S1 for 4_1}
\end{align}
The two-loop invariants $S^{(\textrm{conj})}_2(N)$ are
\begin{align}
&\textrm{Im}[S^{(\textrm{conj})}_2(N)]  \;  \quad  \textrm{for } N=2,\ldots 20 \nn
\\
&=\{0.0882063, 0.289984, 0.618779, 1.13059, 1.89451, 2.96776, 4.40130, \
6.24658,    \nn
\\
&\qquad 8.55519, 11.3786, 14.7680, 18.7749, 23.4506, 28.8465, 35.0139, 42.0042, 49.8689, \nn
\\
&\qquad
58.6593, 68.4268 \} \;.\nn
\end{align}
and their third-difference sequence is
\begin{align}
&\textrm{Im}[S_2^{'''}(N)] \quad  \textrm{for } N=2,\ldots 17 \nn
\\
&= \{  0.0560005, 0.0690888, 0.0572193, 0.0509708, 0.0514399, 0.0516042,0.0513983,  \nn
\\
&\qquad 0.0513494, 0.0513577, 0.0513623, 0.0513673,0.0513741,0.0513805,0.0513860, \nn
\\
&\qquad
0.0513907,0.0513947 \} \;. \nn
\end{align}
The sequence rapidly approaches the value $0.0514...$ which is very close to the number  $\frac{1}{(2\pi)^2}\textrm{vol}(S^3 \backslash \mathbf{4}_1) =0.0514175...$, see  Figure \ref{fig:Re[S1] and Im[S2]}. Thus we numerically confirm that
 \begin{align}
  \textrm{Im}[S^{(\textrm{conj})}_2 (N)] =  \frac{N^3}{24 \pi^2} \textrm{vol}(S^3\backslash \mathbf{4}_1)+(\textrm{sub-leading in $1/N$}) \quad \textrm{as } N\rightarrow \infty \;. \label{S2 for 4_1}
 \end{align}
The three-loop invariants are
 \begin{align}
 &S_3^{(\textrm{conj})}(N) \quad  \textrm{for } N=2,\ldots 9 \nn
 \\
 &= \{-0.0185185, -0.0362503,-0.0425853,-0.0396434,-0.0348546,-0.0312819,  \nn
 \\
 &\qquad -0.0284423, -0.0260191\}\;. \nn
 \end{align}
The first term equals $-1/54$ and matches the result in \cite{Dimofte:2012qj}. Although it is difficult to figure out the leading behavior of $S_3$ at large $N$ from this data, it seems very likely that
 \begin{align}
 \lim_{N\rightarrow \infty}\frac{S_3^{(\textrm{conj})}(N)}{N^3} =0 \label{S3 for 4_1}
 \end{align}
The results \eqref{S0 for 4_1}, \eqref{S1 for 4_1}, \eqref{S2 for 4_1}, \eqref{S3 for 4_1} together confirm the conjecture  \eqref{conjecture} for $M= S^3\backslash \mathbf{4}_1$  up to $n=3$.

\subsubsection{Other knot complements}
Explicit expressions for internal vertices and meridian variables in  the $N$-decomposition of various knot complements in terms of the octahedra's vertex variables are available in the recent version of {\tt SnapPy} up to $N=15$. From these information, it is straightforward to obtain generalized Neunmann-Zagier datum $\{ A_N, B_N,X^{(\textrm{conj})} f_N, f'_N,f''_N\}$ and calculate the perturbative invariants $S^{(\textrm{conj})}_n (N)$.
For five examples of knot complements ($M=S^3\backslash \CK$, $\CK=5_2, 6_1, 6_2,6_3, 7_3$), we have computed $S^{(\textrm{conj})}_{1,2}(N)$ up to $N=12 \sim 15$.
To read off the leading $N^3$-term of these invariants, we plot its third difference sequences in $N$; see Figure \ref{S1,S2 for 5 knots}.
\begin{figure}[h!]
\begin{center}
   \includegraphics[width=.48\textwidth]{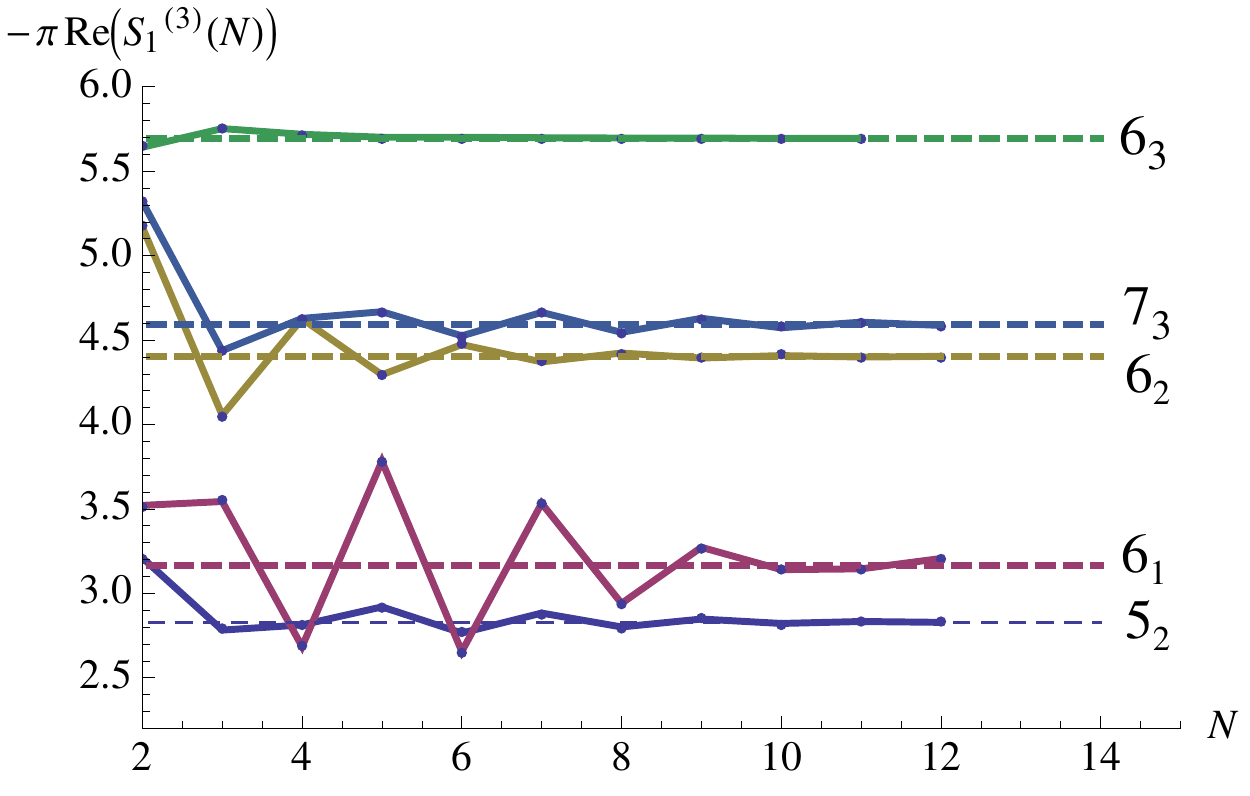}
    \includegraphics[width=.47\textwidth]{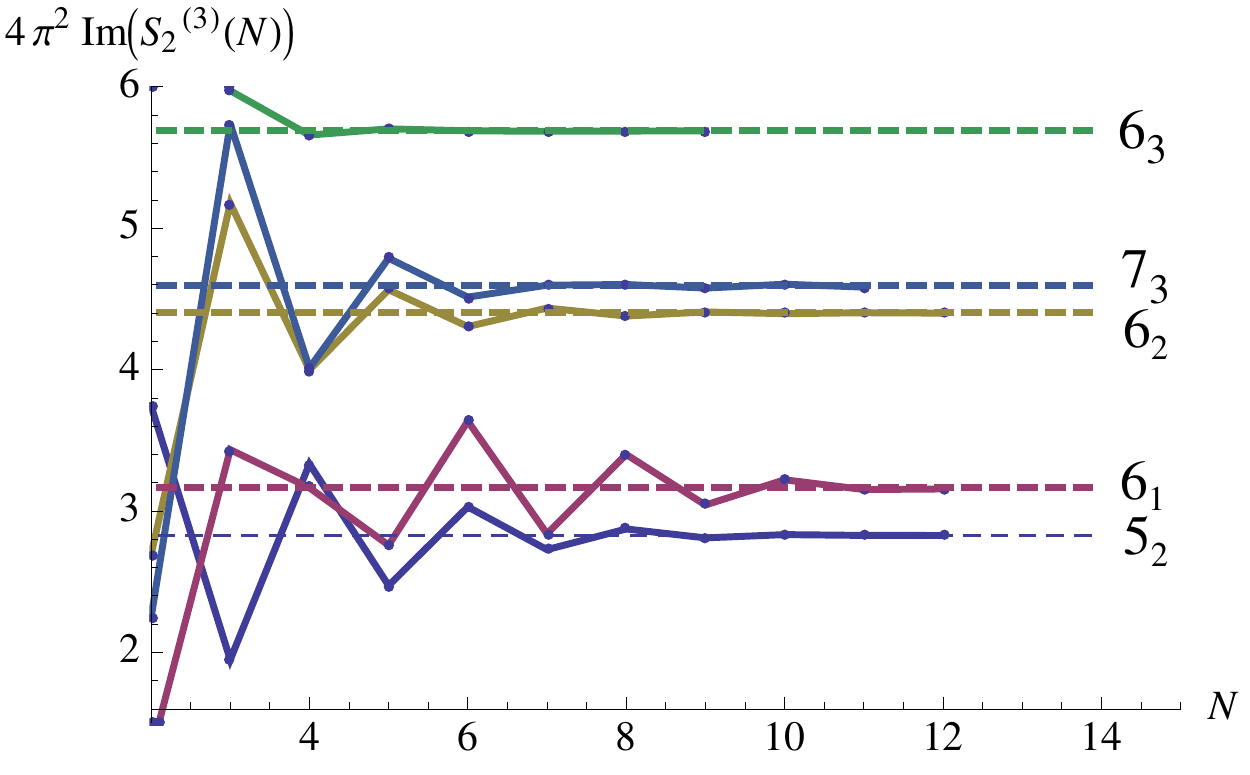}
   \end{center}
   \caption{As $N$ grows, $-\pi {\rm Re}[S_1^{'''}(N)]$ and $4\pi^2{\rm Im}[S_2^{'''}(N)]$ for various knot complements $M=S^3\backslash \CK$ 
   quickly approach $\textrm{vol}(M)$ (dashed line). This phenomenon serves as a numerical evidence for the conjecture \eqref{conjecture} at $n=1, 2$ for $\CK =  5_2,6_1, 6_2, 6_3, 7_3 $.}
    \label{S1,S2 for 5 knots}
\end{figure}
%

%\newpage

\section{Discussion} \label{sec : conclusion}

 We have studied the large $N$ behavior of 3d $T_{N}[M]$ theory  by computing the free energy on a squashed 3-sphere $S^3_b$. 
 The computation has been done indirectly either by using holography (section \ref{sec : sugra}) or by using the 3d-3d correspondence (section \ref{sec : field theory}). We have obtained strong evidences, partly analytic and partly numerical, for perfect agreement of the two results. However, both calculations come with some caveats.  
In the holographic computation, the supergravity solution in \cite{Gauntlett:2000ng} is strictly valid when $M$ is compact,  
and should be modified when $M$ is a knot (or link) complement. 
We assumed that the subtle modification due to the cusp boundary of $M$ would not affect the leading $N^3$-term of the free energy. 
In a related context, the leading $N$-dependence of the theory of class $S$ 
was shown to be insensitive to punctures on Riemann surfaces  \cite{Gaiotto:2009gz}.  In the computation using the 3d-3d correspondence, on the other hand, we relied on two non-trivial assumptions, \eqref{Two things on CS ptn-1}  and \eqref{Two things on CS ptn-2}, to arrive at our main conjecture \eqref{conjecture} on the perturbative expansion of the $PGL(N)$ CS theory.   
Although we have given strong evidences using the state-integral, 
it would be desirable to find alternative, independent ways to 
verify these assumptions and the main conjecture. 

Our analysis can be extended  by adding defects into the system as studied in \cite{Bah:2014dsa}. One can consider an M2 brane wrapped on $AdS_2 \times \gamma$, where $\gamma$ is a one-cycle in $M$. These defects correspond to  line defects in the $T_{N}[M]$ theory. There are two types of supersymmetric line operators $\CW$ and $\widetilde{\CW}$ for generic $b$, located along the  curves at $z=0$ and $w=0$  in $S^3_b$ \eqref{squashed sphere}.  In the 3d-3d correspondence, these line operators might correspond to Wilson loop operators along $\gamma$ in $PGL(N)$ CS theory  on $M$. Wilson loops constructed from holomorphic gauge field $\CA$ and anti-holomorphic gauge field $\bar{\CA}$ correspond to  line operators $\CW$ and $\widetilde{\CW}$, respectively.  Using the gravity solution in section \ref{sec : sugra}, one can  holographically  determine the dependence of the Wilson loop expectation values on $N$ and $\gamma$ at large $N$,
\begin{align}
&\log |\langle  \CW (\gamma;N)\rangle_b | \propto N  \times\ell(\gamma) \times (1+b^2)\;,  \nn
\\ 
&\log |\langle \widetilde{\CW} (\gamma;N)\rangle_b| \propto N  \times \ell(\gamma) \times (1+b^{-2}) \;,
\end{align}
 where $\ell(\gamma)$ denotes the hyperbolic length of $\gamma$. The $b$-dependence was studied in \cite{Farquet:2014bda}. 
Again, the dependence on $\hbar :=2\pi b^2$ is interesting;  it predicts that at large $N$ the Wilson loop expectation values in $PGL(N)$ CS theory are captured by classical and one-loop calculations. The dependence $N \times \ell(\gamma)$ in the classical ($\hbar^{0}$) term can be easily understood in terms of $PGL(N)$ CS theory. The classical part is nothing but the the Wilson loops evaluated at the saddle point $\CA^{(\textrm{conj})}$, which give 
 \begin{align}
 \langle \CW (\gamma;N)\rangle_{\hbar^{0}} &= \textrm{Tr}\; \textrm{Hol}_{\gamma} (\CA^{(\textrm{conj})}) = \textrm{Tr} \;\rho_N \big{(}\textrm{diag}(e^{\ell (\gamma)/2+i\delta (\gamma) }, e^{-\ell(\gamma)/2- i\delta (\gamma)}) \big{)}\;, \nn
\\
 & =  \exp \big{(}(N-1) (\ell/2 +i \delta)\big{)}+\ldots+\exp\big{(}-(N-1) (\ell /2+i \delta)\big{)}\;.
  \end{align}
Here $e^{i \delta(\gamma)}$ is a phase factor. The imaginary part of $\CA^{(\textrm{conj})}_{N=2}$ is constructed using vielbein $e$ and  integration of the flat connection along $\gamma$ gives  a holomony whose eigenvalues are $e^{\pm \frac{1}2\ell(\gamma)}$  up to a phase factor. The first term is dominant at  large $N$ and it explains the $N\times \ell(\gamma)$ behavior in the $\hbar^{0}$ order.  It would be nice if one can check the $N\times \ell(\gamma)$ behavior in the $\hbar^{1}$ order from a direct one-loop computation of Wilson loop expectation values in $PGL(N)$ CS theory.

\vskip 1cm

\begin{acknowledgments}
We are grateful to Kimyeong Lee, Piljin Yi, Jinseok Cho, Seonhwa Kim, Akinori Tanaka, Roland van der Veen, Jun Murakami and  Satoshi Yamaguchi for helpful discussions. DG thanks the organizers of ``Exact Results in SUSY Gauge Theories in Various Dimensions'' at CERN, and also CERN-Korea Theory Collaboration funded by National Research Foundation (Korea), for the hospitality and support. 
\end{acknowledgments}

\newpage

\appendix
\centerline{\large\bf Appendix}

\section{Upward flows in  state-integral} \label{sec : flow analysis}
\begin{figure}[h!]
\begin{center}
   \includegraphics[width=.45\textwidth]{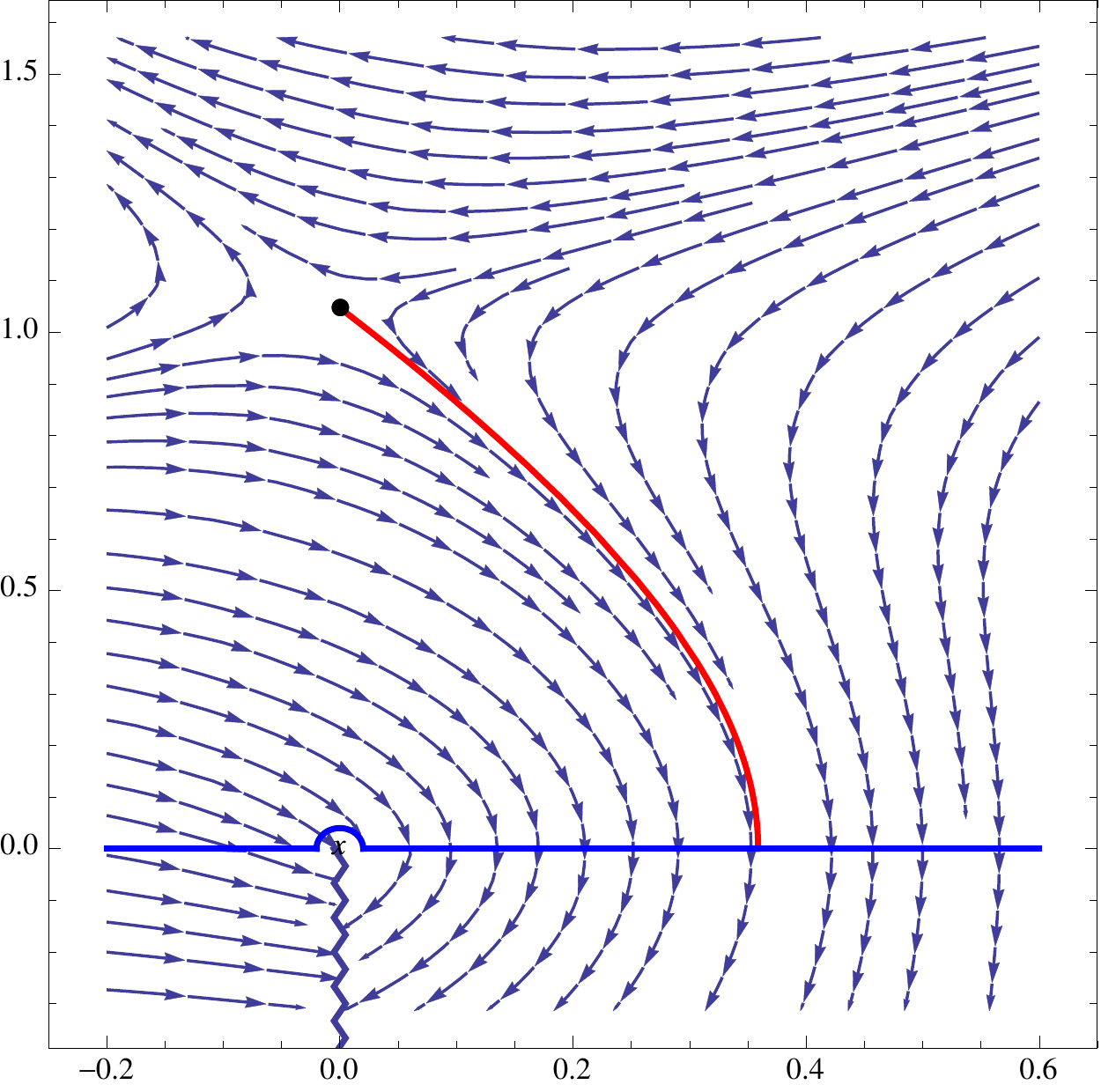}
   \end{center}
   \caption{An upward flow (red line) from a saddle point $(X^{(\textrm{conj})},Y^{(\textrm{conj})})$ (black dot) to $\mathcal{C}_{\mathbb{R}}$ (blue line).  The flow $(X(t),Y(t))$ is located on the  hyperplane $X=Y$ and the figure shows the curve $X(t)$ on a complex plane $\mathbb{C}$. The zigzag line represents the branch cut for $\widetilde{\textrm{Li}_1} (e^{-X})$. }
    \label{4-1gradientflow}
\end{figure}

\begin{figure}[h!]
\begin{center}
   \includegraphics[width=.48\textwidth]{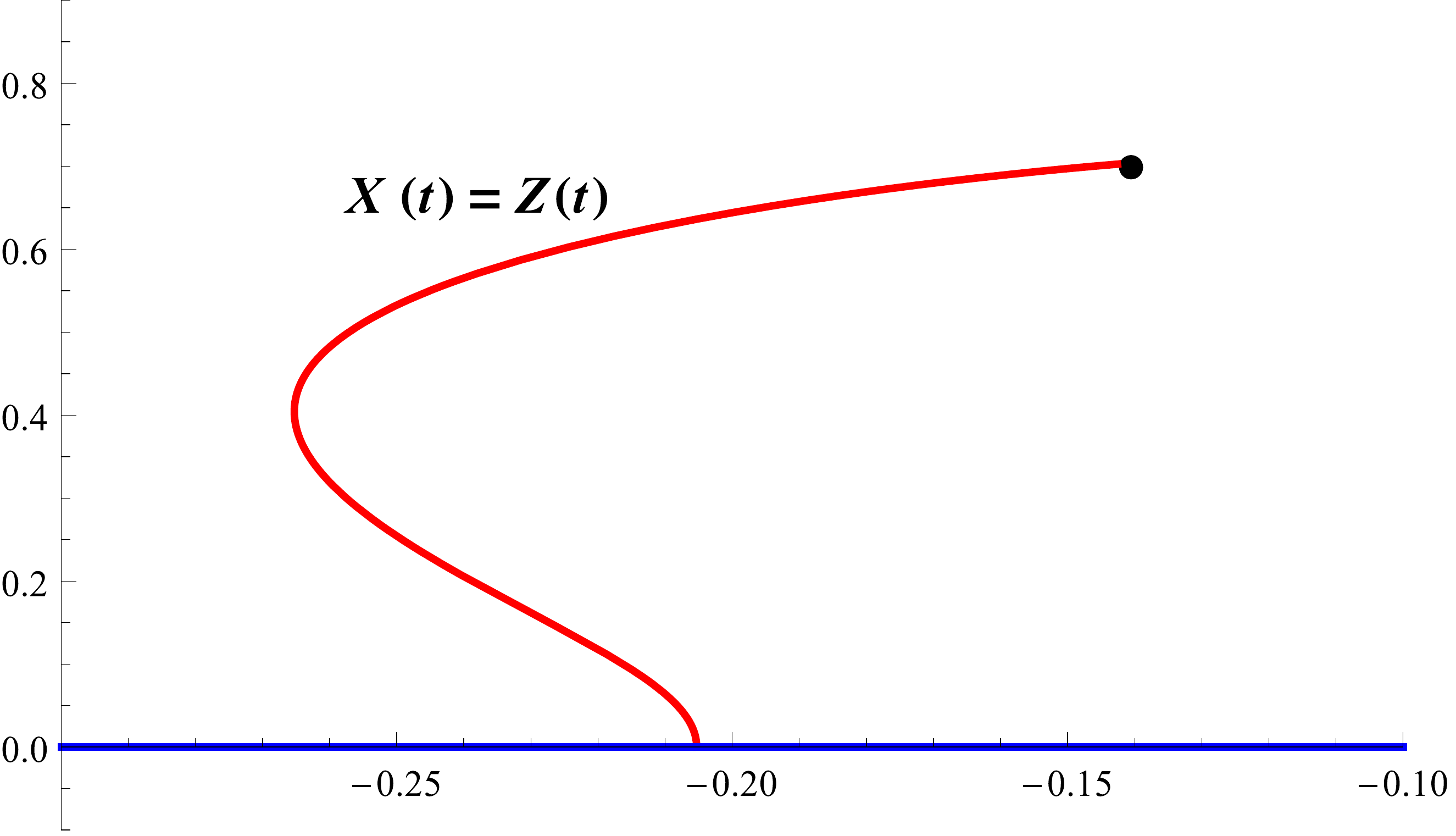}
     \includegraphics[width=.48\textwidth]{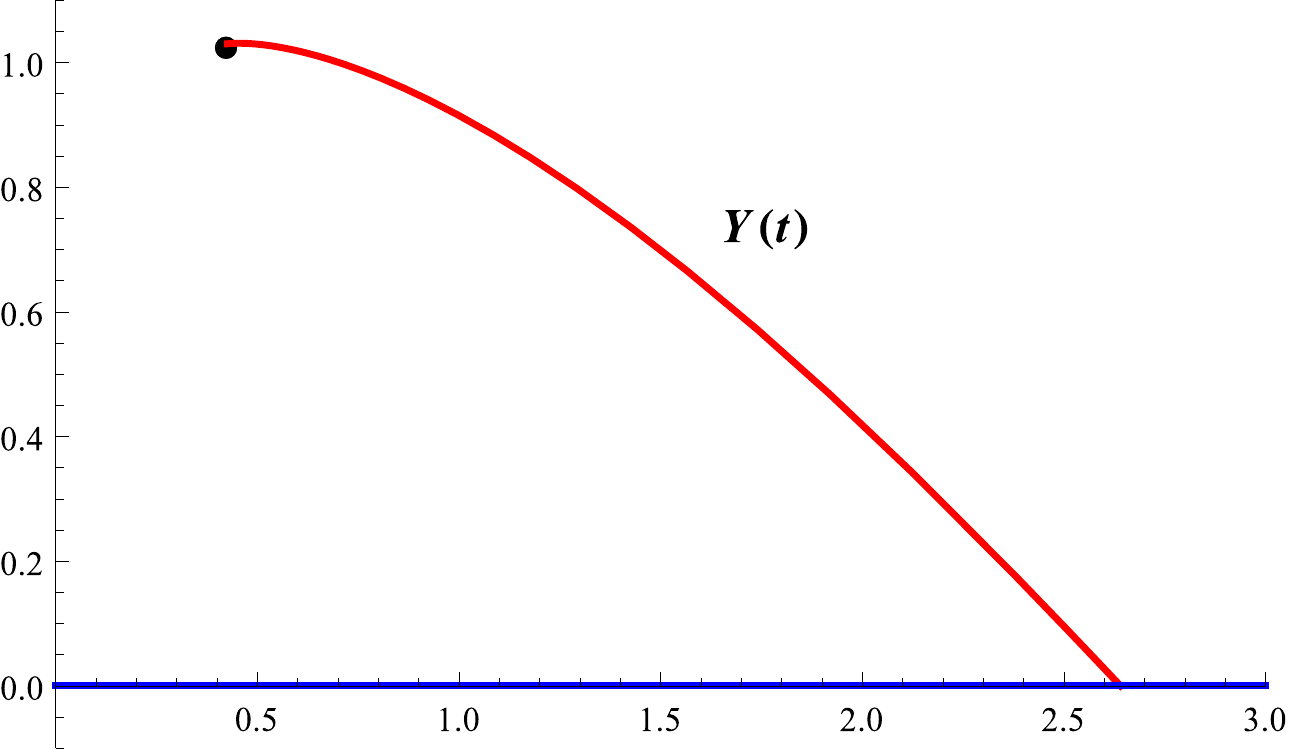}
   \end{center}
   \caption{An upward flow (red line) from a saddle point $(X^{(\textrm{conj})}_{k=0},Y^{(\textrm{conj})}_{k=0},Z^{(\textrm{conj})}_{k=0})$ (black dot) to $\mathcal{C}_{\mathbb{R}}$ (blue line).  The flow $(X(t),Y(t),Z(t))$ are located on the hyperplane $X=Z$ and the left graph and right graph show $X(t)=Z(t)$ and  $Z(t)$, respectively on a complex plane $\mathbb{C}$. }
    \label{5-2gradientflow}
\end{figure}

When $M$ is the figure-eight knot complement ($S^3\backslash \mathbf{4}_1$) and $N=2$, the state-integral is 
\begin{align}
Z^{\textrm{CS}}_{N=2}(\hbar;M) = \int_{\mathcal{C}_{\mathbb{R}}} \frac{dX dY}{2\pi i \hbar}\exp \left( -\frac{1}{\hbar} X Y\right) \psi_\hbar (X) \psi_\hbar (Y)\;.
\end{align}
In the limit $\hbar = 2\pi i b^2 \rightarrow 0$ with real $b$, the logarithm  of the integrand is 
\begin{align}
\CI(X,Y) &\simeq \frac{1}{2\pi i b^2} \big{(}-XY+\widetilde{\textrm{Li}}_2 (e^{-X})+\widetilde{\textrm{Li}}_2 (e^{-Y})\big{)}+o(b^0)
\nn \\
&:= \frac{1}{2 \pi ib^2}\widetilde{\CW}(X,Y)+o(b^0) \;.
\end{align}
There is only one `classical' saddle point  $(X^{(\textrm{conj})}, Y^{(\textrm{conj})})$ satisfying $\partial_X \widetilde{\CW} = \partial_Y \widetilde{\CW}=0$,
\begin{align}
X^{(\textrm{conj})} = Y^{(\textrm{conj})} = i \pi/3\;.
\end{align}
There exists an upward flow from the saddle point to $\mathcal{C}_{\mathbb{R}}$ as shown in the Fig \ref{4-1gradientflow}.
\\
When $M$ is $S^3\backslash \mathbf{5}_2$, the 3-manifold can be triangulated using three tetrahedra. For $N=2$,  the twisted potential is
\begin{align}
\widetilde{\CW}= &-i \pi (X+Y+Z)+\frac{1}2 \big{(}X^2+Y^2+Z^2+2 XY+2YZ\big{)} 
\nn \\
&+\widetilde{\textrm{Li}}_2 (e^{-X})+\widetilde{\textrm{Li}}_2 (e^{-Y})+\widetilde{\textrm{Li}}_2 (e^{-Z})\;.
\end{align}
There are two classes of classical saddle points, 
\begin{align}
&(X,Y,Z)_{k}^{(\textrm{conj})} = (-0.1406+0.703858 i , 0.421799+1.03002 i, -0.1406+0.703858 i )  \nn
\\
&\qquad \qquad \qquad \qquad +2\pi i(k,0,-k) \;, \quad k \in \mathbb{Z}\nn
\\
&(X,Y,Z)_{k}^{(\textrm{geom})} =(\overline{X},\overline{Y},\overline{Z})^{(\textrm{conj})}\;.
\end{align}
Perturbative invariants $S_n$ around each saddle points does not depend on $k$. 
There is no upward flow starting from  $(X,Y,Z)^{(\textrm{geom})}$ to $\mathcal{C}_{\mathbb{R}}$ since $\textrm{Re}[\CI(X^{(\textrm{conj})},Y^{(\textrm{conj})},Z^{(\textrm{conj})}]) = 2.82812=\textrm{vol}(S^3\backslash \mathbf{5}_2)$ is greater than $\textrm{Sup}_{(X,Y,Z)\in \mathcal{C}_{\mathbb{R}}}\textrm{Re}[\CI(X,Y,Z)]=0$. Recall that $\textrm{Re}[\CI]$ never decreases along the upward flow. On the other hand, there is an upward flow from $(X,Y,Z)^{(\textrm{conj})}_{k=0}$ to $\mathcal{C}_{\mathbb{R}}$, which is depicted in Fig \ref{5-2gradientflow}.

\newpage

\bibliography{w3c}

\end{document}